\documentclass[5p,authoryear,fleqn]{elsarticle}
\makeatletter 
\def\ps@pprintTitle{%
 \let\@oddhead\@empty
 \let\@evenhead\@empty
 \let\@evenfoot\@oddfoot} 
\makeatother
\usepackage[utf8]{inputenc} 
\usepackage[T1]{fontenc}
\usepackage[french, english]{babel}
\usepackage{bm}
\usepackage{upgreek}
\usepackage{mathtools}
\usepackage[table]{xcolor}
\usepackage{siunitx}
\usepackage[babel=true]{csquotes} 
\usepackage{amsthm} 
\usepackage{booktabs} 
\usepackage{multirow} 
\usepackage{amssymb} 
\usepackage{float}
\usepackage{hyperref} 
\usepackage[french]{cleveref} 
\newcommand\figurewidth{9 cm}
\begin{document}

\begin{frontmatter}

\title{Learning-based Linear Inversion for Quantitative Pulse-Echo Speed-of-Sound Imaging}

\author{Parisa Salemi Yolgunlu, Jules Blom, Naiara {Korta Martiartu},  and Michael Jaeger}


\begin{abstract}
Computed ultrasound tomography in echo mode generates maps of tissue speed of sound (SoS) from the shift of echoes when detected under varying steering angles by solving a constrained linearized inverse problem. Gradient regularization (GR) allows one to derive a pre-computable linear pseudo-inverse for computationally efficient SoS reconstruction, however, SoS estimates are found to be biased depending on tissue layer geometry. Here, we propose an alternative approach where a learned linear operator minimizes SoS errors on average over a large number of random tissue models that sample the distribution of geometries and SoS values expected in vivo. In an extensive simulation study on liver imaging, we demonstrate that biases are strongly reduced, with residual biases being the result of a partial non-linearity in the actual physical problem. This approach can either be applied directly to echo-shift data (learned pseudo-inverse) or to the SoS maps estimated with GR (learned correction), where the former shows slightly better performance, but the latter requires less extensive training. Physical phantom results confirm the transferability of our results to real ultrasound data. 

\end{abstract}

\begin{keyword}
Handheld ultrasound \sep quantitative ultrasound \sep inverse problem solution \sep pseudo-inverse \sep sound velocity \sep liver imaging
\end{keyword}

\end{frontmatter}

\footnotetext{This work was supported by the Swiss National Science Foundation under project No. 205320-179038. (Corresponding author: Michael Jaeger.)

P. Salemi Yolgunlu and M. Jaeger are with the Institute of Applied Physics, University of Bern, CH-3012 Bern, Switzerland (parisasalemi1991@gmail.com; michael.jaeger@unibe.ch)
Naiara {Korta Martiartu} is with the Signal Processing Laboratory 5 at Ecole polytechnique fédérale de Lausanne, 1015 Lausanne (e-mail: naiara.korta@espci.fr) 
J. Blom is with the Vrije Universiteit Amsterdam, 1081 HV Amsterdam, The Netherlands (e-mail: j.blom@vu.nl)}

\section{Introduction}\label{introduction}
{P}{ulse}-echo speed-of-sound (SoS) imaging is a quantitative technique that leverages the non-invasiveness, the low-cost, and the portability inherent in conventional ultrasound (US) imaging. 
The SoS in tissue is a promising biomarker for breast cancer risk assessment and diagnosis~\cite{li2009vivo,wiskin2012non,sandhu2015frequency,ruby2019breast,Duric2007breast,Flynn2017risk} and quantification of liver disease~\cite{boozari2010evaluation,imbault2017robust,burgio2019ultrasonic,telichko2022noninvasive,stahli2023first} for improved patient care. 
While several techniques have been developed for providing spatially resolved maps of SoS using handheld US ~\cite{robinson1982registration,kondo1990crossed,hesse2013fullwave,jakovljevic2018local,sanabria2018spatial,podkowa2020convolutional,Feigin2020deeplearning,Ali2022layeredMedia,heriard2023refraction,simson2024investigating,Beuret2024,Ali2023ambiguity}, here we focus on computed ultrasound tomography in echo mode (CUTE)~\cite{jaeger2015computed}. This method is based on insonifying the tissue and recording echoes under various combinations of transmit (Tx) and receive (Rx) steering angles~\cite{stahli2020improved}. A mismatch between the anticipated and actual SoS results in an echo localization offset that varies with varying Tx and Rx angles, leading to an echo shift (ES) from which the spatial distribution of SoS can be reconstructed by solving an inverse problem. Despite the limited steering angle range (one-sided detection), axial variations can be detected due to the availability of echoes---and thus of ES data---inside the tissue, however, with a comparably low sensitivity. To stabilize the inversion in the presence of noise and model errors, \textit{a priori} constraints are needed. 
The typical approach to imposing such constraints is by regularizing the solution. Regularization of the L2 norm of the first-order spatial derivatives (gradient regularization - GR) imposes a smoothness constraint in agreement with the assumption that tissue SoS does not vary abruptly~\cite{stahli2020improved}. Together with a linearized forward model, this GR allows the solution of the inverse problem via a linear pseudo-inverse operator. This pseudo-inverse does not depend on individual data samples and can thus be pre-computed. That way, the reconstruction of SoS maps becomes computationally very efficient---an important requirement for real-time imaging. 
Quantitative imaging has been demonstrated in simple phantoms~\cite{stahli2020improved, jaeger2022pulse}, and a clinical pilot study~\cite{stahli2023first} confirmed the promise of CUTE for diagnosing liver steatosis. Nonetheless, the quantitative reproducibility of GR-SoS maps \textit{in vivo} is limited. Stähli et al.~\cite{stahli2020bayesian} observed that liver SoS estimates can substantially vary across different locations on the abdominal wall. They hypothesized that this variation is caused by ES noise and proposed a Bayesian approach: by incorporating an explicit prior on the spatial distribution of the different tissues, the repeatability of liver SoS was substantially improved, from about $\pm 20$~m/s to about $\pm 10$~m/s. Unfortunately, the Bayesian approach is not computationally efficient since it requires a segmentation of B-mode images and re-computing the pseudo-inverse operator for individual data samples. Also, accurate segmentation can be challenging depending on the tissue architecture. An alternative approach was suggested in~\cite{salemi2023excluding} to mitigate the impact of ES noise by excluding noisy areas in the SoS reconstruction process. 

Here we show in a simulation study of liver imaging that---even in noise-free cases---liver SoS values obtained with GR are biased depending on the geometry of the tissue layers of the abdominal wall. We further demonstrate that these biases can be strongly reduced under a different SoS reconstruction paradigm: instead of regularizing the inverse problem solution, the pseudo-inverse itself is constrained with the aim to minimize SoS mean reconstruction errors over a set of SoS-ES data pairs representing the probabilistic distribution of "reasonable" tissue models. This is either implemented as learning a matrix operator that corrects the GR-based pseudo-inverse, henceforth termed "learned correction" (LC), or by directly learning an optimum pseudo-inverse, henceforth termed "learned inverse" (LI). Note that this is different from a recently proposed approach \cite{bezek2024learningforwardmodel} where the forward model is learned, but the solution is obtained with a regularized iterative inversion. Very importantly, the LI and LC matrices are still pre-computable allowing a direct inversion with the mentioned benefit for real-time imaging. 

Our results show that this simple change in paradigm has important and unexpected consequences: when ES data are modeled using the straight-ray approximation, the LC/LI approaches yield accurate liver SoS values independent of layer geometry. This proves that the biases observed with GR are not an imperative result of the CUTE inverse problem being ill-posed. Rather, they are related to the smoothness constraint, which does not represent the sample distribution well. We then perform full simulations of the CUTE image formation process, which includes US wave propagation, signal detection, beamforming, and ES tracking. The learning-based approaches still reduces biases to a great extent, but residual biases remain due to nonlinearities in the SoS-ES relationship. Finally, phantom results demonstrate the transferability of the simulation-based LC and LI to real data. 

This study extends our conference proceeding on the same topic \cite{jaeger2023proceeding} in various aspects: (i) We focus on the direct solution for the learning-based approach as opposed to an iterative solution. (ii) The learning-based approach is presented in a more general way, investigating a learned inversion (LI) in addition to a learned correction (LC). (iii) We provide a thorough investigation of the dependence of performance on training set size. (iv) For the latter, the simulation data was extended from 2000 to 56000 samples. (v) We provide a thorough quantitative comparison of LC/LI with GR. (vi) We confirm the bias reduction performance in a quantitative analysis of physical phantom measurements.

\section{Methods}
\label{sec:method}
In this section, we first review the principle of CUTE (Section \ref{sec:cute_principle}) and highlight the gradient regularization (GR) approach (Section \ref{sec:tikhonov_reg}). Then, we present the learning-based approaches (Section \ref{sec:matrix_reg}) and describe the simulation and processing details (Sections \ref{sec:echo_simulation} and \ref{sec:details}).

\subsection{CUTE principle}
\label{sec:cute_principle}
We refer the reader to~\cite{stahli2020improved} for details on the CUTE image generation process. We summarize here the basics that are needed to understand the difference between the GR and the learning-based approaches. Implementation details used in this study follow further below in sub-section II.D.

The process starts by beamforming the analytic signals (either provided by the machine directly or after Hilbert transform) using delay-and-sum, to create complex-valued radio-frequency (i.e., before envelope detection)(CRF)-mode images for different combinations of transmit (Tx) 
steering angles $\phi_n$ and receive (Rx) steering angles $\varphi_m$ that are chosen from the same range of monotonically increasing equidistant angles indexed by $n$ and $m$. Throughout the beamforming process, we assume a uniform SoS $c_0$.  
Subsequently, we measure the spatially resolved ES between Tx/Rx steering angle pairs having identical mid-angles, where the mid-angle is defined as $0.5(\phi+\varphi)$. The ES at each location is quantified by taking the phase of the complex-valued zero-lag local cross-correlation. This results in ES maps $\theta(x,z,n,m)$, where $n$ and $m$ refer to the Tx angle step from $\phi_n$ to $\phi_{n+1}$ and the Rx angle step from $\varphi_{m+1}$ to $\varphi_m$, respectively. Under the straight-ray approximation, ES data can be related to line integrals of the slowness deviation (SD) $\sigma(x,z)=1/c-\sigma_0$, where $c$ is the true SoS. This forward model is linear and 
can be written in matrix notation as
\begin{equation}
\bm{\uptheta}=\mathbf{M}\bm{\upsigma},
\label{eq_forwardmodel}
\end{equation}
where $\mathbf{M}$ is built from path integration weights that link the vectorized discrete maps of SD, $\bm{\upsigma}$, to the vectorized discrete maps of ES, $\bm{\uptheta}$. These weights are defined assuming that the $\bm{\upsigma}$ between the discrete Cartesian nodes can be obtained by bilinear interpolation. 
The inverse problem related to~\eqref{eq_forwardmodel} is ill-posed and requires regularization.

\subsection{Gradient regularization}
\label{sec:tikhonov_reg}
Our standard CUTE approach \cite{stahli2020improved} uses GR to derive $\bm{\upsigma}$ by solving the least-squares optimization problem
\begin{equation}
\begin{split}
\widehat{\bm{\upsigma}} &= \arg\min_{\bm{\upsigma}}\left\{ \|\mathbf{M}\bm{\upsigma}- \bm{\uptheta}\|^2_{2} \right.\\
&\quad \left. +\lambda^2_x\|\mathbf{D}_{x}\bm{\upsigma}\|^2_{2} +\lambda^2_z\|\mathbf{D}_{z}\bm{\upsigma}\|^2_{2}   \right\}.
\end{split}
\label{eq2}
\end{equation}
The first term is the sum-square data misfit.
The second and third terms stabilize the inversion by constraining the spatial gradient of $\widehat{\bm{\upsigma}}$. Here, $\mathbf{D}_{x}$ and $\mathbf{D}_{z}$ are the finite-difference operators in $x$ and $z$ directions, respectively, and the regularization parameters $\lambda^2_x$ and $\lambda^2_z$ control the smoothness along each direction independently. 
The regularized optimization problem in~\eqref{eq2} has the closed-form solution 
\begin{equation}
\widehat{\bm{\upsigma}} = 
\mathbf{M}^\dag \bm{\uptheta},
\label{eq_slowness}
\end{equation}
where the pseudo-inverse operator $\mathbf{M}^\dag$ is 
\begin{equation}
\mathbf{M}^\dag=\left( \mathbf{M}^{T}  \mathbf{M} +\lambda^2_{x}\mathbf{D}^{T}_{x} \mathbf{D}_{x} + \lambda^2_{z}\mathbf{D}^{T}_{z} \mathbf{D}_{z}\right)^{-1}  \mathbf{M}^{T}. 
\label{eq_pseudo inverse}
\end{equation}
The superscript $T$ stands for the matrix transpose operation. The main advantage of the linear approach in~\eqref{eq_slowness} is that it allows real-time SoS imaging: the operator $\mathbf{M}^\dag$ is independent of ES data; thus, it only needs to be derived once. Then, the multiplication of $\mathbf{M}^\dag$ with ES data can be repeated for successive acquisitions in real time. 

Finally, the SoS map is obtained from the SD map as
\begin{equation}
c(x,z) = \frac{1}{\widehat{\sigma}(x,z)+\sigma_0}. 
\label{eq_slowness to SoS}
\end{equation}
\subsection{Learning-based reconstruction}
\label{sec:matrix_reg}
The smoothness constraint in~\eqref{eq2} effectively stabilizes the inverse problem; however, in  Section~\ref{sec results:tikhonov regularization}, we show that it leads to layer geometry-dependent biases. 
In this study, we propose an alternative paradigm to SoS reconstruction. We seek for a solution of the form
\begin{equation}
\widehat{\widehat{\bm{\upsigma}}} = \mathbf{\Gamma}\mathbf{y},
\label{eq4}
\end{equation}
where the estimated SD $\widehat{\widehat{\bm{\upsigma}}}$ is derived from measurement data $\mathbf{y}$ via a matrix multiplication, allowing real-time reconstructions.
$\mathbf{y}$ can be either the ES data $\bm{\uptheta}$ or the SD maps $\widehat{\bm{\upsigma}}$ reconstructed by the GR approach. Note that---in the first case---the matrix $\mathbf{\Gamma}$ takes the role of the pseudo-inverse of $\mathbf{M}$ (learned inverse, LI) . In the second case, it acts as a correction to the $\widehat{\bm{\upsigma}}$ derived from $\bm{\uptheta}$ via $\mathbf{M}^\dag$ (learned correction, LC). Our goal is to find an instance $\mathbf{\Gamma}_\mathrm{opt}$ of $\mathbf{\Gamma}$ that minimizes a loss function representing the average reconstruction error over a set of ground truth SD tissue models that sample the probabilistic distribution of expected tissue geometries and SoS values:
\begin{equation}
L(\mathbf{\Gamma}) \coloneq \frac{1}{N N_x N_z}\|\bm{\Sigma} - \widehat{\widehat{\bm{\Sigma}}} \|^2_{2}.
\label{eq_loss}
\end{equation}
Here, $N$ is the number of tissue models, and $N_x$ and $N_z$ are the target grid dimensions along $x$ and $z$ directions, respectively. The columns of the matrix $\bm{\Sigma}$ are the $\bm{\upsigma}$, and the columns of the matrix $\widehat{\widehat{\bm{\Sigma}}}$ are the $\widehat{\widehat{\bm{\upsigma}}}$, with 
\begin{equation}
\widehat{\widehat{\bm{\Sigma}}} = \mathbf{\Gamma}\mathbf{Y},
\label{eq5}
\end{equation} 
where the columns of the matrix $\mathbf{Y}$ contain the $\mathbf{y}$ (i.e., either $\widehat{\bm{\upsigma}}$ or $\bm{\uptheta}$). 
The prefactor in~\eqref{eq_loss} ensures that loss values are comparable between different choices of $N$ and grid size. 

The optimum $\mathbf{\Gamma}_\mathrm{opt}$ is defined as the solution to the following regularized optimization problem:
\begin{equation}
\mathbf{\Gamma}_\mathrm{opt} = \arg\min_{\mathbf{\Gamma}}\{ L(\mathbf{\Gamma}) +\frac{\gamma^2}{N_x N_z}\|\mathbf{\Gamma}\|^2_{2} \}.
\label{eq_minimumcondition}
\end{equation}

Note that this definition of $\mathbf{\Gamma}_\mathrm{opt}$ corresponds to a multivariate linear regression where the model parameters (matrix weights) are estimated from training samples $\{\mathbf{y},\bm{\upsigma}\}$ ($N$: training set size). It includes a regularization of the square Frobenius norm $\|\mathbf{\Gamma}\|^2_{2}$ with weight parameter $\gamma^2$ (ridge regression). This has a different purpose than in ~\eqref{eq2}: rather than imposing a prior on $\widehat{\widehat{\bm{\upsigma}}}$, it controls the extent of overfitting of the model parameters. Eq.~\eqref{eq_minimumcondition} has the standard closed-form solution: 
\begin{equation}
\mathbf{\Gamma}_\mathrm{opt} = \bm{\Sigma}\mathbf{Y}^{T} 
\left(\mathbf{Y}\mathbf{Y}^{T} + N\gamma^2 \mathbf{I}\right)^{-1},
\label{eq_MR}
\end{equation}
where $\mathbf{I}$ is the identity matrix. The value of $\mathbf{Y}\mathbf{Y}^{T}$ is proportional to the number $N$ of data vectors contained in $\mathbf{Y}$; thus, the relative weight of $N\gamma^2\mathbf{I}$ compared to $\mathbf{Y}\mathbf{Y}^{T}$ does not depend on $N$.  

\subsection{Echo shift data simulation}
\label{sec:echo_simulation}
The derivation of $\mathbf{\Gamma}_\mathrm{opt}$ requires a large number of input-target output pairs $\{\mathbf{y},\bm{\upsigma}\}$ for tissue models sampling the distribution of geometries and SoS values expected \textit{in vivo}.
For this purpose, we use a fast simulation framework. A procedural generator randomly generates SoS models on a 50~mm~$\times$~50~mm Cartesian grid with 0.15~mm grid spacing. Using the hybrid angular spectrum technique~\cite{Vyas2012has}, we simulate Tx and Rx wavefields for individual elements of a linear array probe with 128 elements, 0.3~mm pitch, 5~MHz center frequency, and 4~MHz bandwidth. The interaction of these wavefields with a 2D random reflectivity distribution is modeled using the first-order Born approximation: for each grid point, the scattered field detected by a combination of Tx and Rx elements is the convolution in time of the corresponding wavefields at this grid point, multiplied by its reflectivity. For each Tx/Rx element pair, the total detected scattered wavefield is the sum over all grid points. Based on this full-matrix capture, plane-wave US datasets are synthesized for angles ranging from \ang{-27} to \ang{27}, with a \ang{0.5} step. 

The remainder of the signal processing closely follows the methodology detailed in \cite{stahli2020improved}: 
From the plane-wave data, we generate Tx/Rx steered CRF-mode images for Tx/Rx angles chosen from the set $[\ang{-25}:\ang{2.5}:\ang{25}]$, in an image area of 38.4~mm $(x)$ by 45~mm $(z)$. Tx steering is implemented via plane-wave compounding~\cite{montaldo2009coherent} (weighted sum over the range of plane-wave angles), and Rx steering via k-space filtering, both using an angular aperture radius of \ang{2.5}. 
ES is quantified as the pixel-wise echo phase shift between adjacent angle pairs having the same mid-angle. For this, we compute the zero-lag local cross-correlation as the pixel-wise Hermitian product followed by a convolution with a Hann window functions with 1~mm full-width half-maximum (FWHM) in $x$ and $z$. 
The $\ang{2.5}$ angle step is chosen small to avoid phase aliasing. To keep the number of ES maps small, the angle step is increased to $\ang{10}$ by summing ES data from Tx/Rx angle pairs ($\phi_n$, $\varphi_{m+1}$) to ($\phi_{n+1}$, $\varphi_m$), with $\phi_n$ and $\varphi_m$ from the set $[\ang{-25}:\ang{10}:\ang{25}]$. This results in $25$ ES maps. Taking into account Tx-Rx reciprocity, this number is further reduced to $10$ ES maps. To further reduce ES data, these maps are downsampled to a coarser Cartesian grid with 1~mm grid spacing. 
GR-SD maps are then obtained as outlined in section~\ref{sec:tikhonov_reg}. The size of the vectorized ES ($\bm{\uptheta}$) and SD ($\widehat{\bm{\upsigma}}$) is $45\times38\times10$~=~$17100$ and $45\times38$~=~$1710$ elements, respectively. For display, SoS maps are computed by Eq.~\ref{eq_slowness to SoS} and cut to a range of 40~mm in $z$. 

With the above methodology, the simulation of ES data for one individual tissue model takes 40~s, allowing us to obtain 1000 simulations in 11 hours (Matlab R2021a, Nvidia GeForce RTX 3070, AMD Ryzen 9 5950X). These simulations will in the follwing be referred to as \enquote{wave model.} To investigate to what extent the observed biases are inherent in the linear forward/inverse model, we also generate ES data using the linear forward model in~\eqref{eq_forwardmodel}. To mitigate the inverse crime, ES data is simulated with the grid resolution of the CRF-mode images and then downsampled to the resolution used in the wave model. This model will in the following be referred to as \enquote{linear model.} 

\begin{figure}[!t]
\centerline{\includegraphics[width=\figurewidth]{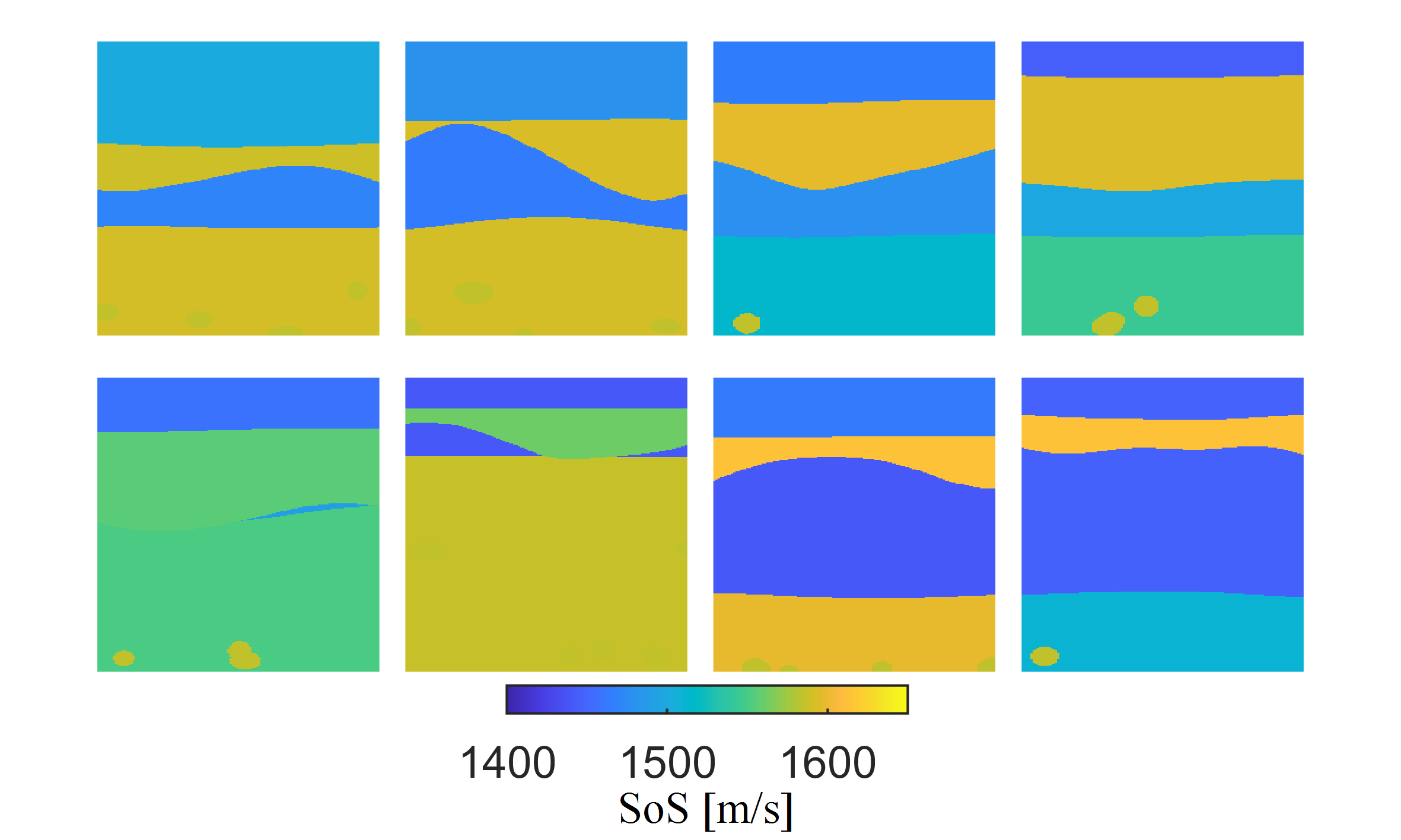}}
\caption{Representative selection of SoS models created by the procedural tissue model random generator. 
Each panel shows an area of 40~mm~$\times$~40~mm corresponding to the SoS map area used in the remainder of the article. 
}
\label{fig: SoS models}
\end{figure}

For this study---to mimic liver imaging---the procedural random generator was designed to create SoS models with piece-wise constant SoS mimicking the different abdominal tissues: subcutaneous fat (F1), muscle (M), pre-peritoneal fat (F2), liver (L), and blood vessels. The generator first places horizontal layer interfaces at random depths. The interfaces are then randomly distorted to create layers with laterally varying diameters, in some cases leading to the absence of tissues in areas where the distorted interfaces overlap. The SoS values of the different tissues are then randomly chosen from uniform distributions within predefined value ranges. These ranges are summarized in Table~\ref{tab1}, together with the most important parameters used for CUTE processing. Fig.~\ref{fig: SoS models} illustrates the breadth of the obtained variety of SoS models.

\begin{table}[!t]
\centering
\caption{Parameter summary}
\label{tab1}
\setlength{\tabcolsep}{8pt}
\begin{tabular}{|p{115pt}|p{100pt}|}
\hline
\multicolumn{2}{|c|}{\bf{Tissue model}} \\
\hline
grid resolution ($x$, $z$) & \qty{0.15}{\mm}, \qty{0.15}{\mm}\\
\hline
fat SoS & \qtyrange{1440}{1500}{\m\per\s} \\
\hline
muscle SoS & \qtyrange{1555}{1615}{\m\per\s} \\
\hline
liver SoS & \qtyrange{1500}{1600}{\m\per\s} \\
\hline
blood SoS & \qty{1585}{\m\per\s} \\
\hline
\multicolumn{2}{|c|}{\bf{CUTE processing}} \\
\hline
beamforming SoS & \qty{1540}{\m\per\s} \\
\hline
CRF-mode grid resolution ($x$, $z$) & \qty{0.3}{\mm}, \qty{0.15}{\mm} \\
\hline 
CRF-mode grid size ($x$, $z$) & \qty{128}{}, \qty{300}{} \\
\hline
ES Tx/Rx steering angles & \qtyrange{-25}{25}{^\circ} in \qty{10}{^\circ} steps \\
\hline 
ES and SD grid resolution ($x$, $z$) & \qty{1}{\mm}, \qty{1}{\mm}  \\
\hline
ES and SD grid size ($x$, $z$) & \qty{38}{}, \qty{45}{}  \\
\hline
unit of SD $\sigma$ & \unit{\um\per\mm} \\
\hline
unit of ES $\theta$ & \unit{\us} \\
\hline
unit of $\mathbf{M}$ &  \unit{\mm}   \\
\hline 
unit of $\lambda^2$ & \unit{\square\mm} \\
\hline
unit of $\gamma^2$, learned correction & \unit{\square\us\per\square\mm} \\
\hline 
unit of $\gamma^2$, learned pseudo-inverse & \unit{\square\us} \\
\hline 
unit of loss value & \unit{\square\us\per\square\mm} \\
\hline 
\end{tabular}
\end{table}

\subsection{Training and Testing Details}
\label{sec:details}
For defining the target output, the ground truth SD maps are resampled to the same grid resolution as the GR-SD maps (1~mm $\times$ 1~mm). The achievable spatial resolution of reconstructed SD maps is typically worse than that, limited by the contrast-resolution trade-off of the CUTE inverse problem. If we use the resampled target SD maps directly, the loss defined in~\eqref{eq_loss} can thus be influenced by the disagreement between the predicted and target resolution at sharp SoS transitions between different layers. To avoid the learning being biased by this disagreement, we generate target SD maps where we mimic a spatial resolution similar to what is typically obtained in a GR approach. This is achieved by convolving the ground truth SD maps---before downsampling---with Hann windows with 1.4~mm FWHM along the $x$ direction and 3.1~mm FWHM along the $z$ direction. Figs.~\ref{figure GR SoS maps}(a) and (b) illustrate the resolution reduction from the tissue models (used for simulation) to the target maps (used for training).

For training and testing of $\mathbf{\Gamma}_\text{opt}$, we generated a set of 28000 pairs $\{\mathbf{y},\bm{\upsigma}\}$. The size of this set is doubled to 56000 pairs by data augmentation: because the imaging problem is symmetric with respect to the $z$-axis, this is achieved by mirroring the SD maps and the ES maps about this axis. In addition, ES maps have to be re-sorted so that the mirrored Tx/Rx angles coincide with the intended angles. The augmented set is split into 50000 training, 5000 validation, and 1000 test samples. 

The validation set is used to investigate the predictive accuracy (validation loss analogue to Eq.~\ref{eq_loss}) of LC/LI on unseen samples, as a function of $\gamma^2$ and training set size $N$. For each $N$, the $\gamma^{2*}$ value leading to the minimum validation loss is used further to analyze the accuracy of layer L's SoS estimates in the test set. Per test sample, we define the \enquote{bias} as the mean signed difference between the reconstructed SoS $\hat{c}(x,z)$ and the ground truth $c(x,z)$ inside the set $\mathbb{L}$ of pixels pertaining to layer L: 
\begin{equation}
bias=\sum\limits_{(x,z)\in\mathbb{L}}{\hat{c}(x,z)-c(x,z)}.
\label{eq: bias definition}
\end{equation}

To investigate how LC/LI influence structural information while reducing bias, we quantify the disagreement between reconstructed and target layer interfaces. For this purpose, we define the \enquote{edge error} as the difference between $\hat{c}(x,z)$ and the ground truth $c(x,z)$. To avoid that this difference includes the bias, we remove low spatial frequencies by subtracting the convolution of the difference with a 2D Hann window kernel $K$ with 7 mm FWHM along the $x$ and $z$ direction, i.e, 
\begin{equation}
edge\;error(x,z) =(\hat{c}(x,z)-c(x,z))*(\delta(x,z)-K(x,z)),
\label{eq: bias definition}
\end{equation} 
where $\delta(x,z)$ is $1$ for $(x=0,z=0)$ and $0$ elsewhere. 
For a statistical analysis, we compute for each test sample the root-mean square (RMS) edge error over all pixels.  

\subsection{Physical data} 
\label{sec:details}
 To test the performance of the proposed method in real physical data, we acquired data from physical twins of a selected number of tissue models (described later), as well as from a healthy volunteer (informed consent obtained, in compliance with the ethical principles of the Declaration of Helsinki 2018). For data acquisition, we used a Vantage~128 research US system (Verasonics Inc., WA, USA) with an L7-4 linear probe (ATL Philips, WA, USA) that has the same number of elements, element pitch, and center frequency as used in the simulations. 

\section{Results}
\label{sec:results}
In this section, we first illustrate the geometry-dependent biases in the GR approach. Then, we demonstrate the effectiveness of the LC and LI approaches in reducing these biases, and investigate the role of the beamforming SoS $c_0$ for our results. Finally, we apply LC and LI to physical phantom data. Throughout this section, the values of
$\lambda_x^2$, $\lambda_z^2$, $\gamma^2$, and loss are written without units to improve readability (see Table~\ref{tab1} for units).
Quantitative metric results are summarized in Table~\ref{tab:metric_table} at the end of the manuscript. 

\subsection{Gradient regularization}
\label{sec results:tikhonov regularization}

\begin{figure}[!t]
\centerline{\includegraphics[width=\figurewidth]{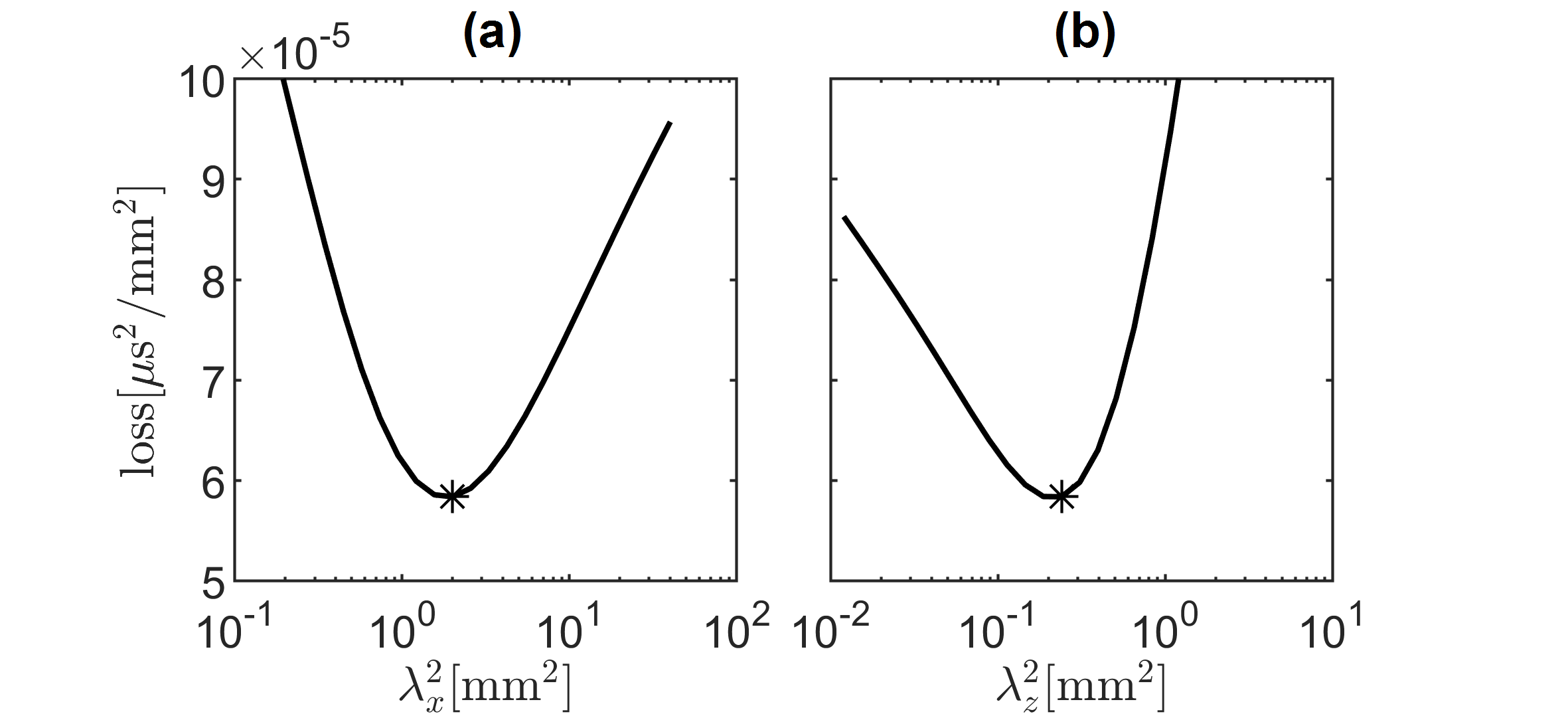}}
\caption{Minimum loss of GR on 6000 samples. (a) Loss as a function of $\lambda^2_x$ for $\lambda^2_z=$~0.24. (b) Loss as a function of $\lambda^2_z$ for $\lambda^2_x=$~2.0. The asterisks indicate the optimum parameters $\lambda^{2*}_x=$~2.0, $\lambda^{2*}_z=$~0.24.}
\label{fig: GR optimisation}
\end{figure}

To ensure a fair comparison between GR and LC/LI, the regularization parameter values in GR are optimized so that $\mathbf{M}^\dag$ minimizes the same loss function~\eqref{eq_loss} as used for LC/LI, over the 5000 validation samples of ES input---simulated with the wave model---and target SD maps. 
The optimum values were found by alternating line searches along the two coordinates ($\lambda^{2}_x$ and $\lambda^{2}_z$), to $\lambda^{2*}_x=\num{2.0}$ and $\lambda^{2*}_z=\num{0.24}$. The minimum loss is $5.9\cdot 10^{-5}$. Fig.~\ref{fig: GR optimisation} depicts the dependence of the loss value on $\lambda^2_x$ and $\lambda^2_z$ around these optimum values.  

\begin{figure}[!t]
\centerline{\includegraphics[width=\figurewidth]{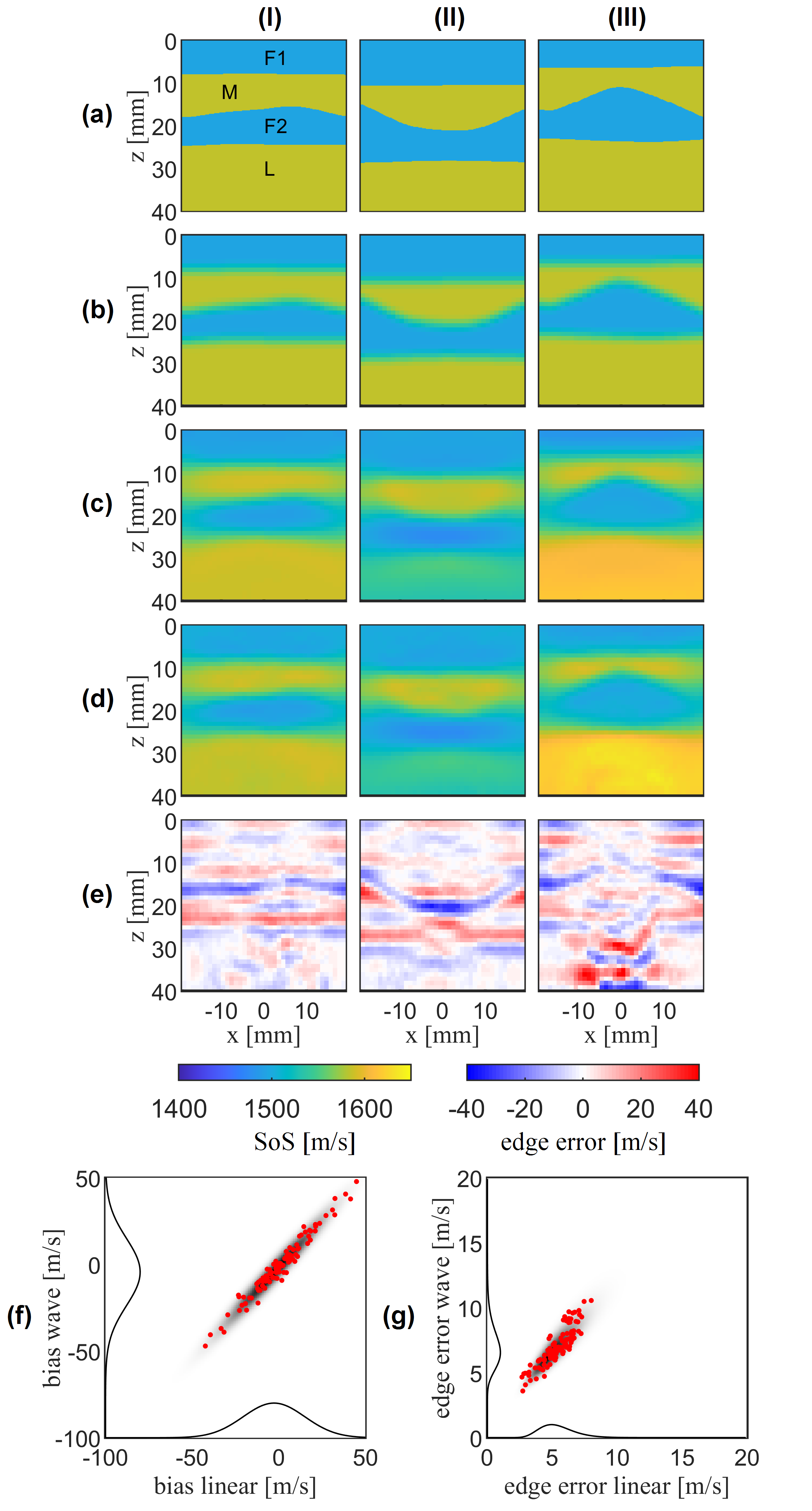}}
\caption{Demonstration of SoS biases in GR-SoS maps, for three example tissue models (I) to (III). (a) Ground truth. Different tissues are marked as sf (subcutaneous fat), m (muscle), pf (pre-peritoneal fat), and l (liver). (b) Target SoS maps. (c)-(d) GR-SoS maps when ES is simulated using the linear and wave model, respectively. (e) Edge error map corresponding to (d). (f) Layer L biases and (g) RMS edge errors for the wave model vs. linear model. Normal distribution fits (log-normal for (g)) are shown for the $1000$ test samples as gray areas together with the values of $100$ examples (red dots). The black profiles are the projections of the fit onto each axis.
}
\label{figure GR SoS maps}
\end{figure}

Utilizing the optimum parameters, we reconstruct the GR-SD maps from ES data derived from the linear and the wave model. 
Fig.~\ref{figure GR SoS maps} illustrates the geometry-dependent SoS biases for three exemplary tissue models, (I) to (III). The reduced-resolution target SoS maps are shown in row~(b). For ease of visual comparison, the same SoS values were chosen (1495~m/s for F1 and F2; 1585~m/s for M and L). While the layer L's reconstructed SoS is roughly accurate for (I), it is visibly underestimated for (II) and overestimated for (III). 
In all cases, this bias is similar for the linear and the wave model. This observation is confirmed in (f), which shows the layer L biases obtained with the linear vs. the wave model for the 1000 test samples. Both models yield a broad range of bias values (with a standard deviation (STD) of $18$~m/s and $19$~m/s, respectively), which are, however, strongly correlated between the two propagation models. This correlation demonstrates that a large part of the observed biases is inherent in the SoS inversion approach rather than in a deviation of the US propagation from the straight-ray assumption. Row (e) shows the 2D resolved edge error for the wave model for later reference. Largest errors are related to the layer interfaces but also to wave-related artifacts as in used for simulation (III) towards the lower image edge. The RMS edge error values for the 1000 test samples in (g) indicate that edge errors are generally stronger for the wave ($7.1\pm1.5$) than for the linear model ($5.4\pm1.2$).  

\subsection{Learned correction---linear model}
\label{sec: matrix reg linear forward}
In a first step, we investigate the learning-based approach as a correction to the GR-SD maps $\widehat{\bm{\upsigma}}$ (LC) and depart from a scenario where $\bm{\uptheta}$ is simulated using the linear model. This allows us to investigate the performance limit of LC in an idealized scenario where the relationship between $\bm{\upsigma}$ and $\bm{\uptheta}$ (and therefore $\widehat{\bm{\upsigma}}$) is perfectly linear. 

\begin{figure}[!t]
\centerline{\includegraphics[width=\figurewidth]{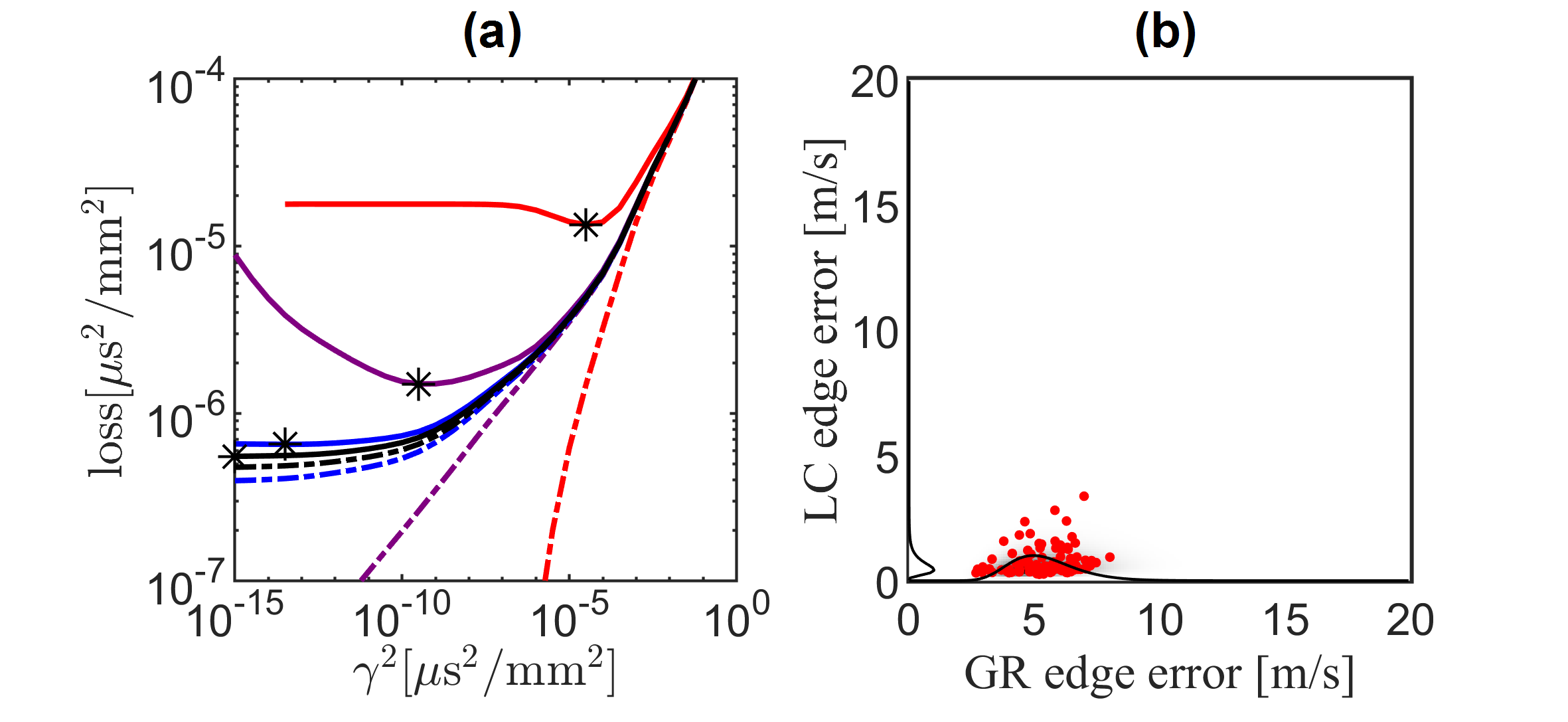}}
\caption{(a) Loss performance of LC when usingthe linear model for simulating ES data. Training (dashed lines) and validation loss (solid lines) as a function of $\gamma^2$ for different $N$ (red: $50$, purple: $1500$, blue: $10000$, black: $50000$). The curve for $N=50$ is not shown for smallest $\gamma^2$ where it is determined by numerical errors. Asterisks denote the validation loss minima. (b) RMS edge errors for LC vs. GR. The log-normal distribution fit (gray area and projections) for $1000$ test samples is shown together with the values for 100 examples (red dots). 
}
\label{fig: losses linear forward model}
\end{figure}

We evaluate LC results for various different training set sizes $N$ ($50$, $1500$, $10000$, and $50000$). Fig.~\ref{fig: losses linear forward model}(a) shows the training and validation loss as a function of $\gamma^2$ for the different $N$. 
In all cases, the training and validation loss curves converge for increasing $\gamma^2$ and diverge for decreasing $\gamma^2$. This behavior is expected as $\gamma^2$ controls overfitting. For any $N$, there exists a $\gamma^2$ above which overfitting is avoided so that the model learned by $\mathbf{\Gamma}_\mathrm{opt}$ generalizes well to the validation set. Smaller $\gamma^2$ result in overfitting so that $\mathbf{\Gamma}_\mathrm{opt}$ performs better on training than on validation data. The larger $N$ is, the closer the validation and training loss curves become, until they nearly coincide for $N=50000$. For increasing $N$, the minimum validation loss and the optimum $\gamma^{2*}$ at which this minimum is achieved decrease.  
The minimum loss values (at corresponding $\gamma^{2*}$ for the different $N$) are: $1.3\cdot10^{-5}$ ($3.1\cdot10^{-5}$, $N=50$), $1.5\cdot10^{-6}$ ($3.1\cdot10^{-10}$, $N=1500$), $6.4\cdot10^{-7}$ ($3.1\cdot10^{-14}$, $N=10000$), and $5.4\cdot10^{-7}$ ($1\cdot10^{-15}$, $N=50000$). Note that already for $N=50$, the minimum loss is below the $5.9 \cdot 10^{-5}$ that was obtained for GR. Note also that it is not substantially reduced from $N=10000$ to $N=50000$ and converges to around $5.1\cdot10^{-7}$. This result indicates that $N=10000$ training samples are sufficient to avoid overfitting.  

\begin{figure}[!t]
\centerline{\includegraphics[width=\figurewidth]{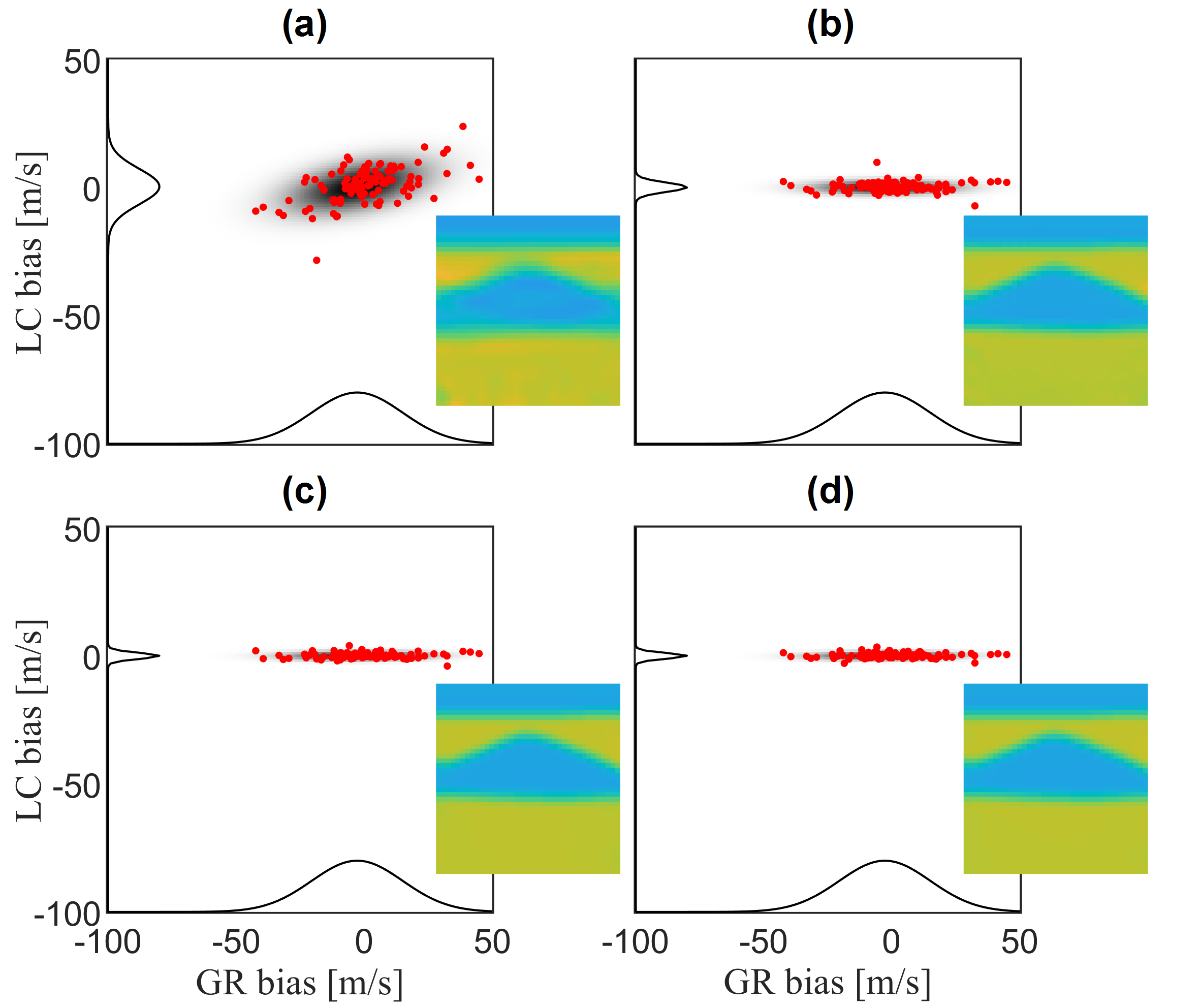}}
\caption{Layer L SoS biases obtained with the LC vs. GR approach when the linear model is used for simulating ES data. LC was optimized for training set sizes $N$ of (a) $50$,  (b) $1500$, (c) $10000$, and (d) $50000$. Gray area and black profiles: normal distribution fit to the 1000 test samples. Red dots: biases for 100 test samples. For each $N$, insets show the LC-SoS maps obtained for the example (III).
}
\label{fig: biases linear forward model}
\end{figure}

After determining $\gamma^{2*}$ for each $N$, we test how the corresponding $\mathbf{\Gamma}_\text{opt}$ performs in terms of layer L SoS biases. 
Figs.~\ref{fig: biases linear forward model}(a)-(d) show, for the different $N$, the biases obtained in the test set ($1000$ samples) with LC vs. GR. While the biases for GR are distributed about the previously mentioned STD of $18$~m/s, the distribution is much narrower for the LC approach: after training on $N=50$ samples, the STD is only $6.9$~m/s, and it decreases to $1.8 $~m/s for $N=1500$ and further to $1.1$~m/s for $N=50000$. Fig.~\ref{fig: biases linear forward model} exemplary shows for tissue model (III) the LC-SoS map when applying the $\mathbf{\Gamma}_\text{opt}$ for the different $N$ to the corresponding GR-SoS map from Fig.~\ref{figure GR SoS maps}(c): from a somewhat blurred appearance (but reduced bias) for $N=50$, it develops into an accurate representation of the target map already for $N=1500$. 

\begin{figure}[!t]
\centerline{\includegraphics[width=\figurewidth]{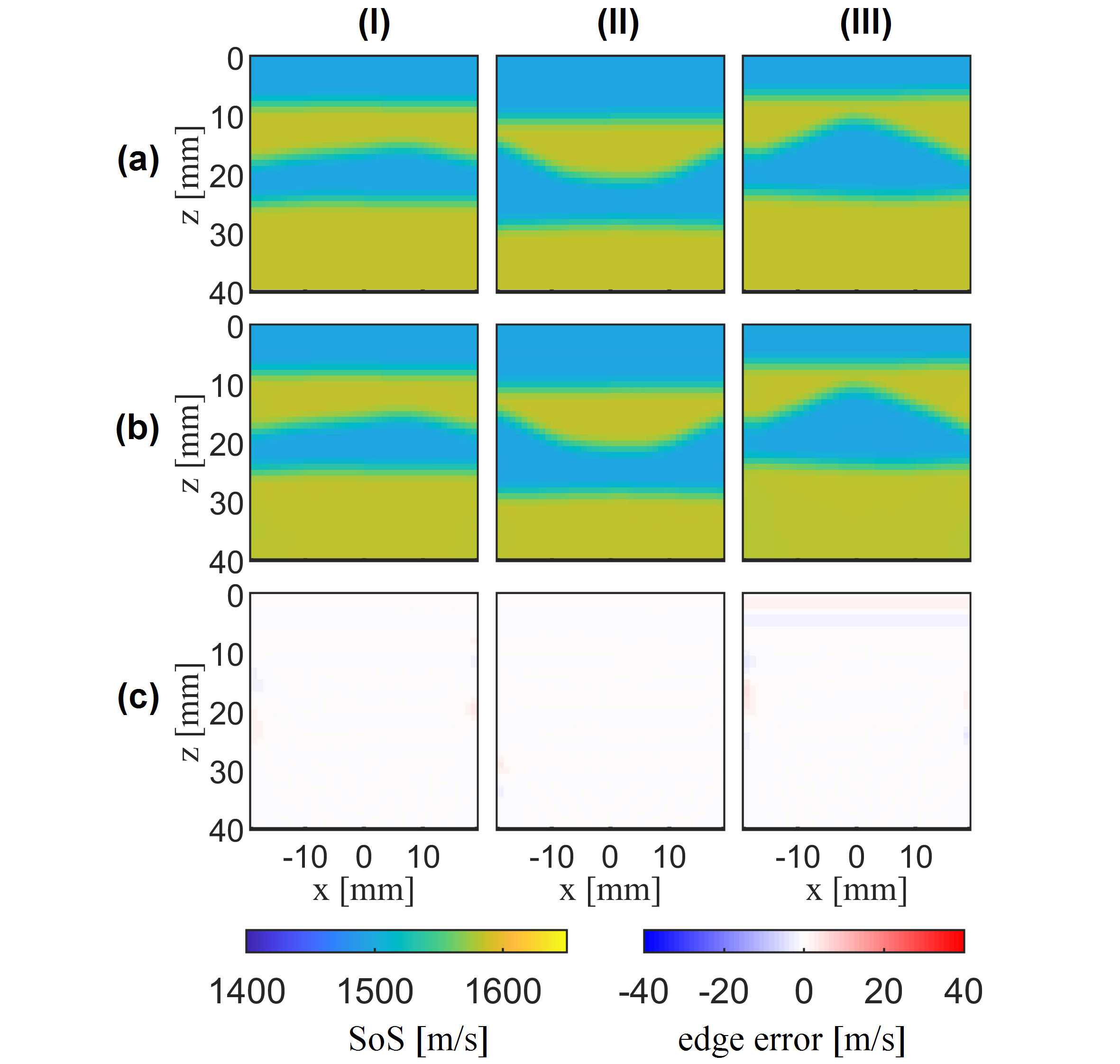}}
\caption{LC results (training set size $N=5000$) for the example tissue models (I) to (III) when using the linear model for ES simulation. (a) Target SoS maps. (b) Corresponding LC-SoS maps.(c) 2D-resolved edge error maps.}
\label{fig: SoS maps linear forward model}
\end{figure}

Fig.~\ref{fig: SoS maps linear forward model}(b) illustrates the SoS map quality obtained with LC for $N=50000$ for the three example tissue models. LC accurately reconstructs the target SoS maps shown for comparison in (a).  The improved accuracy of layer L's SoS compared to Fig.~\ref{figure GR SoS maps}(c) is obtained without loss of edge definition. On the contrary: the  2D edge error maps in (c) are near zero. The improved edge definition compared to GR is confirmed by the RMS edge errors for the 1000 test samples in Fig.~\ref{fig: losses linear forward model}(b): the error distribution is reduced from $5.4\pm1.2$~m/s to $0.68\pm0.55$~m/s.

These results demonstrate that---despite the ill-posed CUTE inverse problem---an accurate reconstruction of liver SoS is, in principle, possible while at the same time improving structural information. The biases observed with GR are therefore not an intrinsic limitation of CUTE nor of the straight ray approximation, but a consequence of the GR strategy. 

\subsection{Learned correction---wave model}
\label{sec: matrix reg full wave forward} 
When the ES is described by the linear forward model, the LC approach can thus provide highly accurate SoS maps. In reality, however, the wave nature of US propagation (diffraction, refraction, and interference) but also echo placement errors introduce a nonlinearity in the relation between SD and ES. Here, we evaluate their influence on the effectiveness of LC in reducing biases, using SD data simulated with the wave model. 

\begin{figure}[!t]
\centerline{\includegraphics[width=\figurewidth]{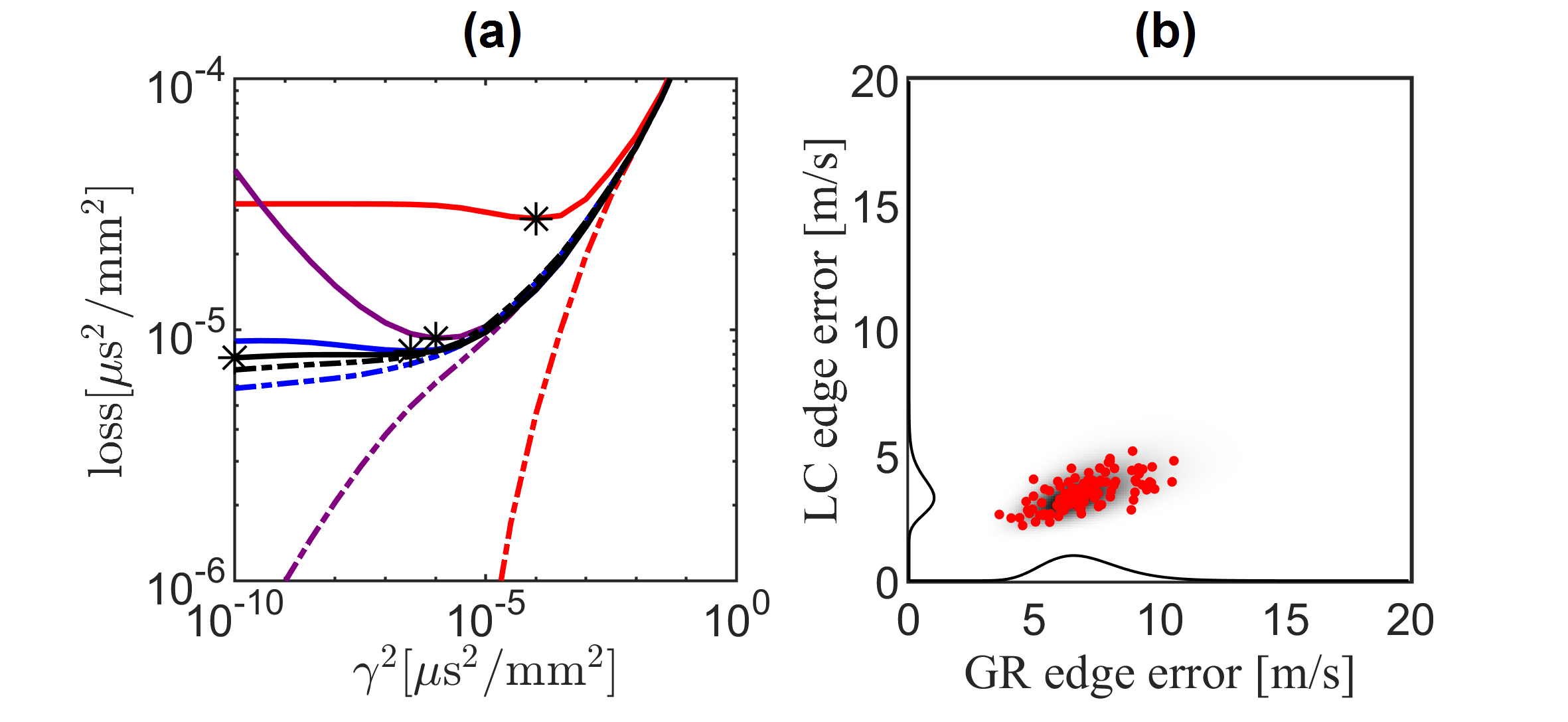}}
\caption{(a) Loss performance of LC when using the wave model for simulating ES data. Training (dashed lines) and validation loss (solid lines) as a function of $\gamma^2$ for different $N$ (red: $50$, purple: $1500$, blue: $10000$, black: $50000$). Asterisks denote the validation loss minima. (b) RMS edge errors for LC vs. GR. The log-normal distribution fit (gray area and projections) for $1000$ test samples is shown together with the values for $100$ test examples (red dots). 
}
\label{fig: losses full wave model}
\end{figure}

Fig.~\ref{fig: losses full wave model}(a) shows training and validation loss as a function of $\gamma^2$ for the different $N$ (note the different scales for loss value and $\gamma^2$ than in Fig.~\ref{fig: losses linear forward model}). The gross behavior of these curves is similar to what was seen for the linear model in Fig.~\ref{fig: losses linear forward model}: For increasing $N$, the minimum loss values and the $\gamma^{2*}$ at which these values are obtained decrease, and for $N=10000$ and $N=50000$ training and test loss curves converge. 
However, the loss reduction is substantially less pronounced, and the minimum loss only slightly decreases for $N\geq1500$. The minimum loss values (at corresponding $\gamma^{2*}$ for the different $N$) are: $2.8\cdot10^{-5}$ ($1.0\cdot10^{-4}$, $N=50$), $9.1\cdot10^{-6}$ ($1.0\cdot10^{-6}$, $N=1500$), $8.1\cdot10^{-6}$ ($3.1\cdot10^{-7}$, $N=10000$), and $7.6\cdot10^{-6}$ ($1\cdot10^{-10}$, $N=50000$).  
The latter value is a factor $7.8$ smaller than the $5.9\cdot10^{-5}$ that was obtained with GR, indicating that LC still reduces the minimum achievable loss compared to GR, even in a scenario with wave propagation effects. 

\begin{figure}[!t]
\centerline{\includegraphics[width=\figurewidth]{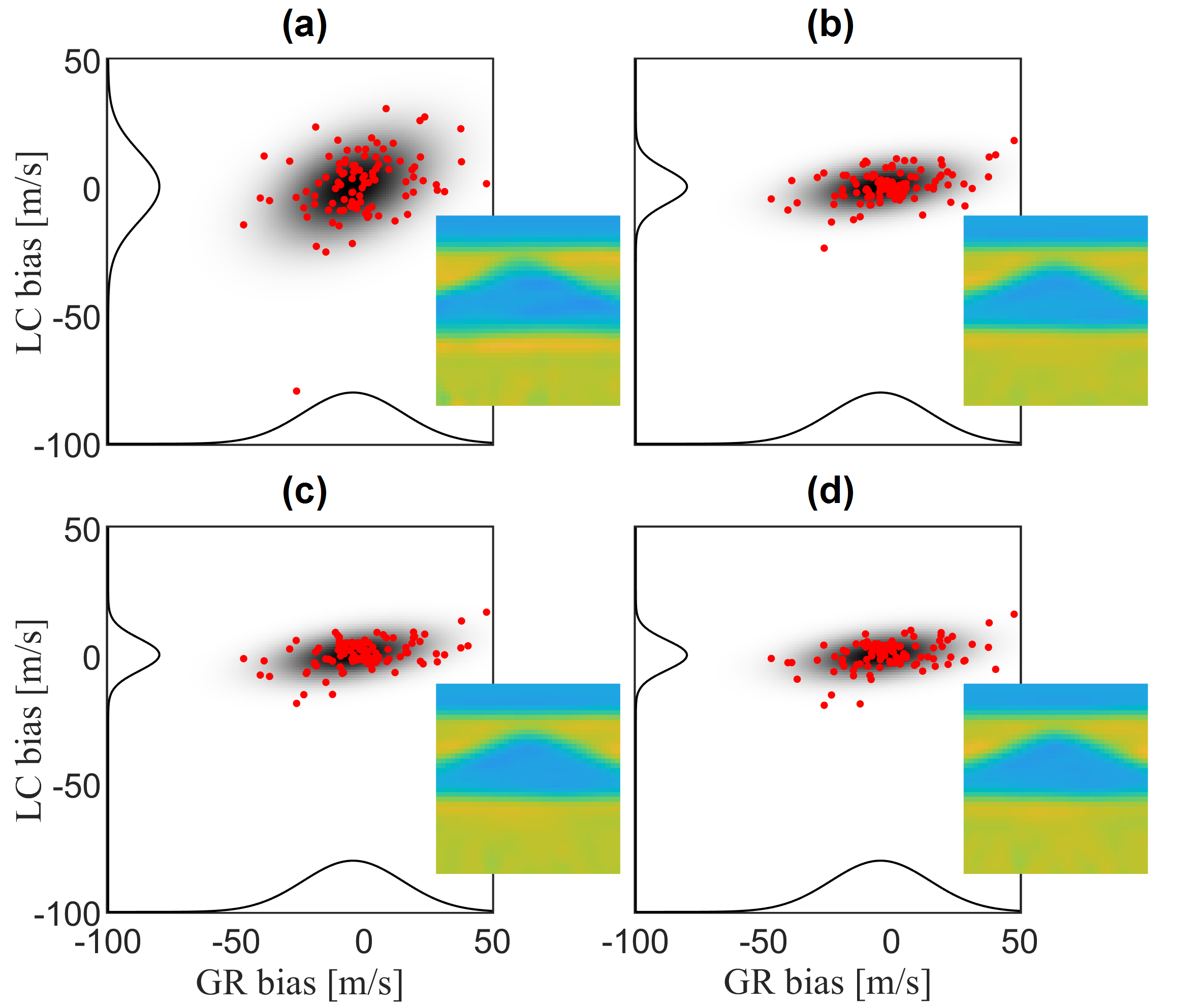}}
\caption{Layer L SoS biases obtained with the LC vs. GR approach when the wave model is used for simulating ES data. LC was optimized for training set sizes $N$ of (a) $50$,  (b) $1500$, (c) $10000$, and (d) $50000$. Gray area and black profiles: normal distribution fit to the 1000 test samples. Red dots: biases for 100 test samples. For each $N$, insets show the LC-SoS maps obtained for the tissue model (III).
} 
\label{fig: biases full wave model}
\end{figure}

The less pronounced reduction of the test loss observed for the wave compared to the linear model is reflected in the SoS biases of layer L: Figs.~\ref{fig: biases full wave model}(a)-(d) show a larger width of the bias distribution obtained with LC than in Fig.~\ref{fig: biases linear forward model}, and this width is only slightly reduced above $N=1500$. The STD values are $14$~m/s ($N=50$), $6.7$~m/s ($N=1500$), $6.3$~m/s ($N=10000$), and $5.9$~m/s ($N=50000$). 
The larger distribution width than for the linear model shows residual biases due to a nonlinear relation between the ground-truth SD and GR-SD maps, introduced by the wave nature and echo placement errors. Despite this, the distribution STD for LC when using $50000$ training samples ($5.9$~m/s) is  $3.2$ times smaller than for GR ($19$~m/s). 
Again, the insets in Fig.~\ref{fig: biases full wave model} allow us to appreciate how the image quality evolves with $N$. As in the linear model case, the quality visibly improves from $N=50$ to $N=1500$ in terms of edge definition and artifact level, but little improvement is seen above $N=1500$. 

\begin{figure}[!t]
\centerline{\includegraphics[width=\figurewidth]{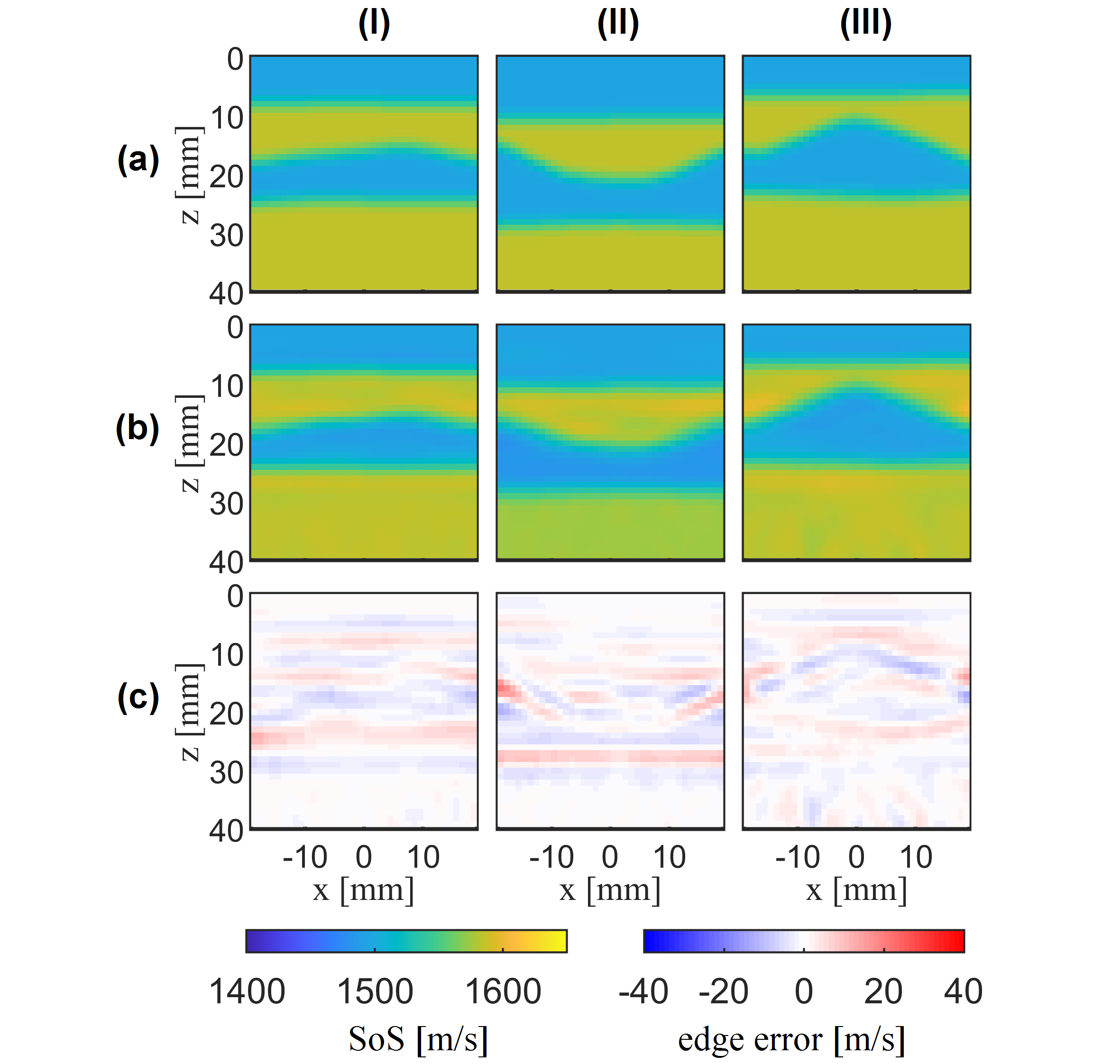}}
\caption{LC results (training set size $N=5000$) for the example tissue models (I) to (III) when using the wave model for ES simulation. (a) Target SoS maps. (b) Corresponding LC-SoS maps.(c) 2D-resolved edge error maps.} 
\label{fig: SoS maps full wave model}
\end{figure} 

The LC-SoS maps of the three example tissue models shown for $N=50000$ in Fig.~\ref{fig: SoS maps full wave model} represent the target SoS maps still well, however, example (II) exemplifies a weak negative bias of layer L. In addition, the 2D-resolved edge error maps in Fig.~\ref{fig: SoS maps full wave model}(c) show an increased edge error level compared to the linear model, however, still below the level observed for GR in Fig.~\ref{figure GR SoS maps}(e). The RMS edge error values in Fig.~\ref{fig: losses full wave model}(b) confirm that the distribution range is reduced, from $7.1\pm1.5$~m/s to $3.5\pm0.78$~m/s. Overall, LC substantially reduces the biases observed for GR while improving structural information. 


\subsection{Learned pseudo-inverse---wave model}
\label{sec: matrix reg echo shift} 
In LC, a matrix operator was trained to transform GR-SD maps into corrected SD maps. Thereby, the pseudo-inverse that generated GR-SD maps from ES data was derived from an explicit forward model. Alternatively, the pseudo-inverse itself can be learned so that it optimally transforms ES data to SD maps, obliterating the need for a forward model. Here, we investigate the performance of this LI approach using ES data simulated with the wave model. 

\begin{figure}[!t]
\centerline{\includegraphics[width=\figurewidth]{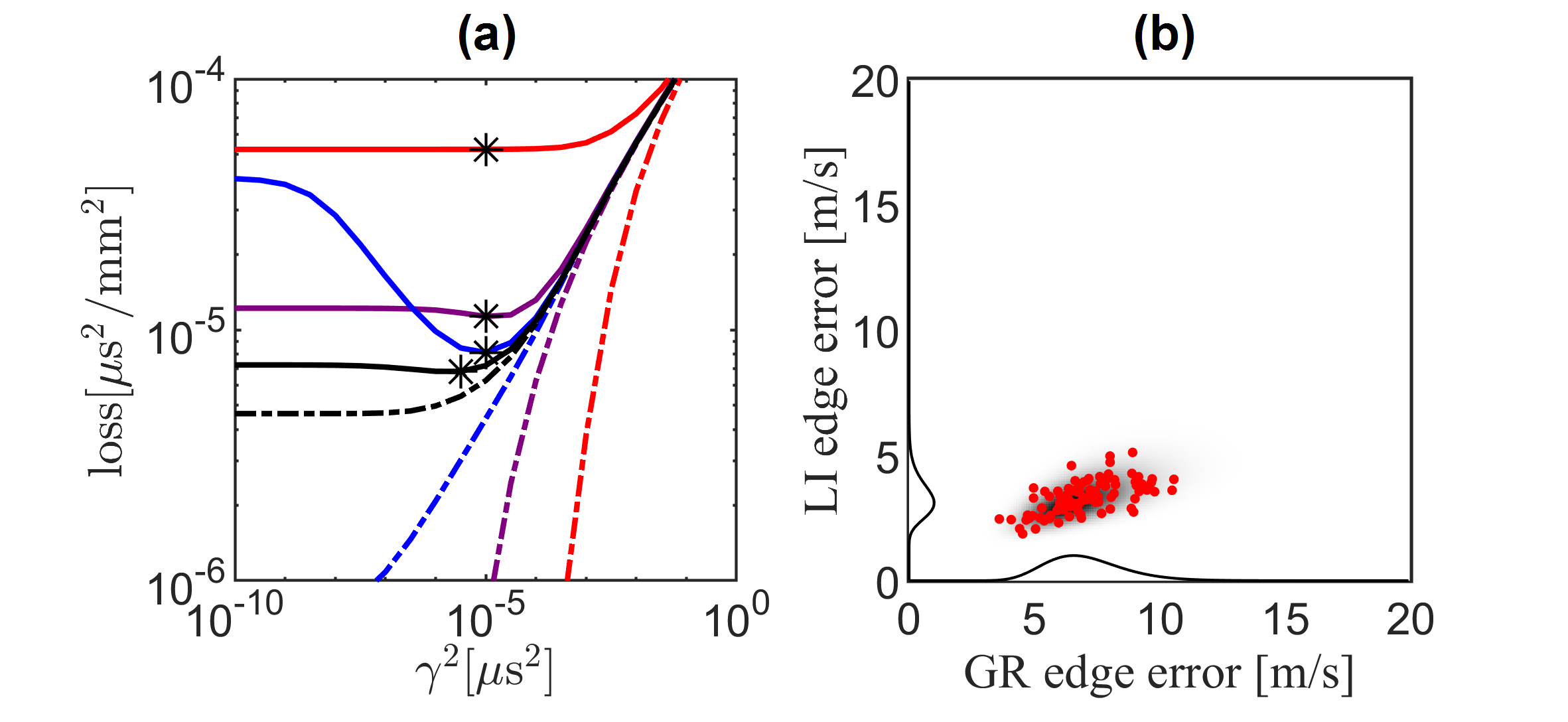}}
\caption{(a) Loss performance of LI when using the wave model for simulating ES data. Training (dashed lines) and validation loss (solid lines) as a function of $\gamma^2$ for different $N$ (red: $50$, purple: $1500$, blue: $10000$, black: $50000$). Asterisks denote the validation loss minima. (b) RMS edge errors for LI vs. GR. The log-normal distribution fit (gray area and projections) for $1000$ test samples is shown together with the values for $100$ test examples (red dots). 
}
\label{fig: losses full wave echo}
\end{figure}

The $\gamma^2$-dependent loss curves for different $N$ are shown in Fig.~\ref{fig: losses full wave echo}. Note that the unit of $\gamma^2$ differs compared to LC because the input data type is different; thus, the values of $\gamma^2$ are not comparable to the ones in LC. 
The minimum validation loss values (at corresponding $\gamma^{2*}$ for the different $N$) are: $5.2\cdot10^{-5}$ ($1.0\cdot10^{-5}$, $N=50$), $1.1\cdot10^{-5}$ ($1.0\cdot10^{-5}$, $N=1500$), $8.0\cdot10^{-6}$ ($1.0\cdot10^{-5}$, $N=10000$), and $6.7\cdot10^{-6}$ ($3.2\cdot10^{-6}$, $N=50000$). While the gross behavior of these curves is similar to the ones observed for LC---the minimum validation loss and the corresponding $\gamma^{2*}$ decrease with increasing $N$---one can note important differences: (i) for $N\leq1500$, the minimum validation loss values are larger than for LC; (ii) the minimum validation loss $6.7\cdot10^{-6}$ for $N=50000$ is smaller than the $8.3\cdot10^{-6}$ observed for LC; and (iii) the training and validation loss for $N=50000$ are further apart and the validation loss exhibits a local minimum. The latter two observations indicate that even larger $N$ can lead to further convergence of training and validation loss, towards an extrapolated loss value of around $6\cdot10^{-6}$. We conclude that, while LI performs worse for small $N$, it can provide higher accuracy than LC but requires more data to fully exploit its potential. 

\begin{figure}[!t]
\centerline{\includegraphics[width=\figurewidth]{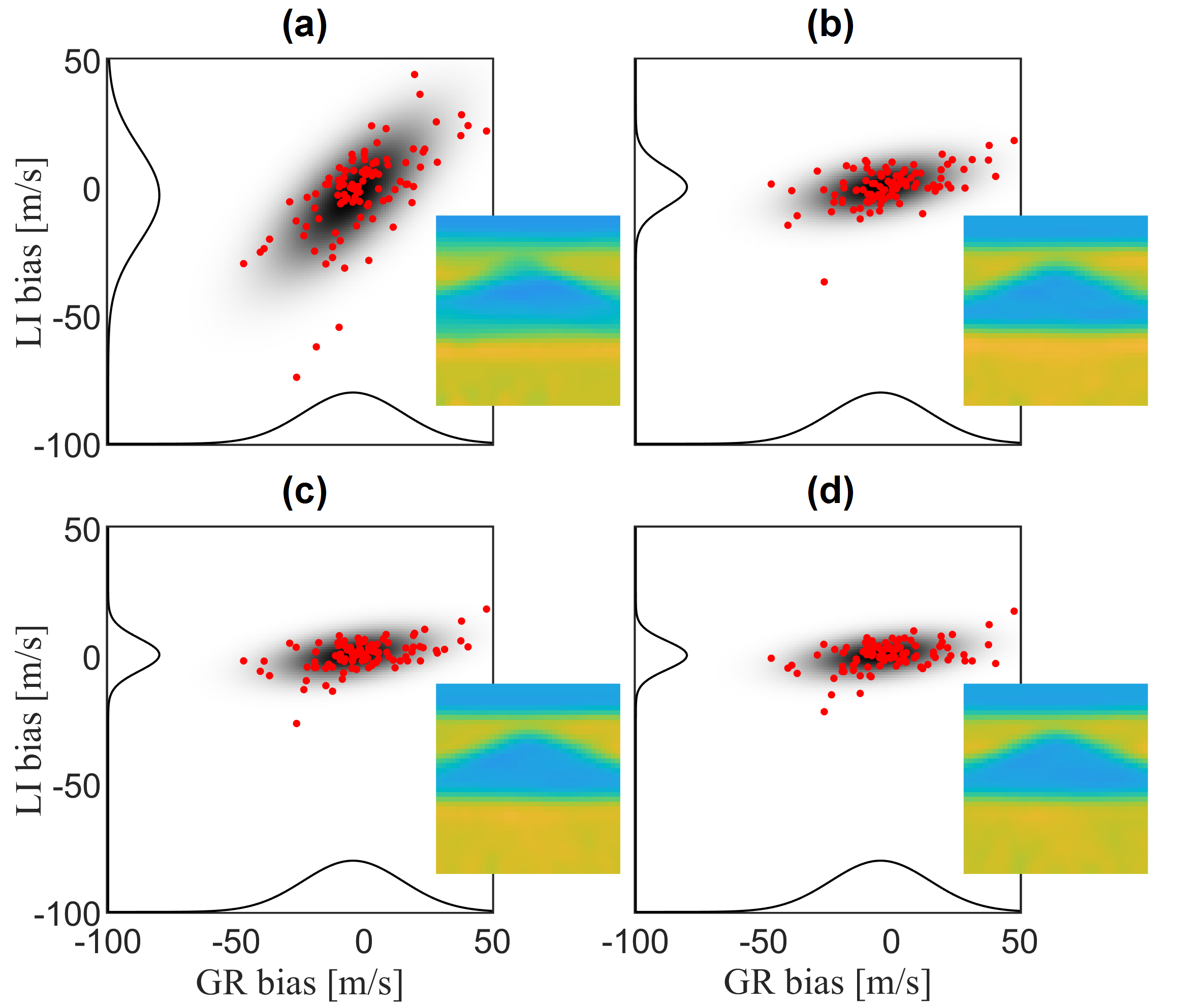}}
\caption{
Layer L SoS biases obtained with the LI vs. GR approach when the wave model is used for simulating ES data. LI was optimized for training set sizes $N$ of (a) 50,  (b) 1500, (c) 10000, and (d) 50000. Gray area and black profiles: normal distribution fit to the 1000 test samples. Red dots: biases for 100 test samples. For each $N$, insets show the LI-SoS maps obtained for the tissue model (III).
} 
\label{fig: biases full wave echo}
\end{figure}

The layer L's SoS bias distributions in Fig.~\ref{fig: biases full wave echo} draw a similar picture: whereas the width of the bias distribution is larger for $N=50$ to $N=10000$ (STD = $19$~m/s, $7.6$~m/, and $6.4$~m/s respectively), it is slightly narrower for $N=50000$ ($5.6$~m/s) than for the same $N$ for LC ($5.9$~m/s). 

\begin{figure}[!t]
\centerline{\includegraphics[width=\figurewidth]{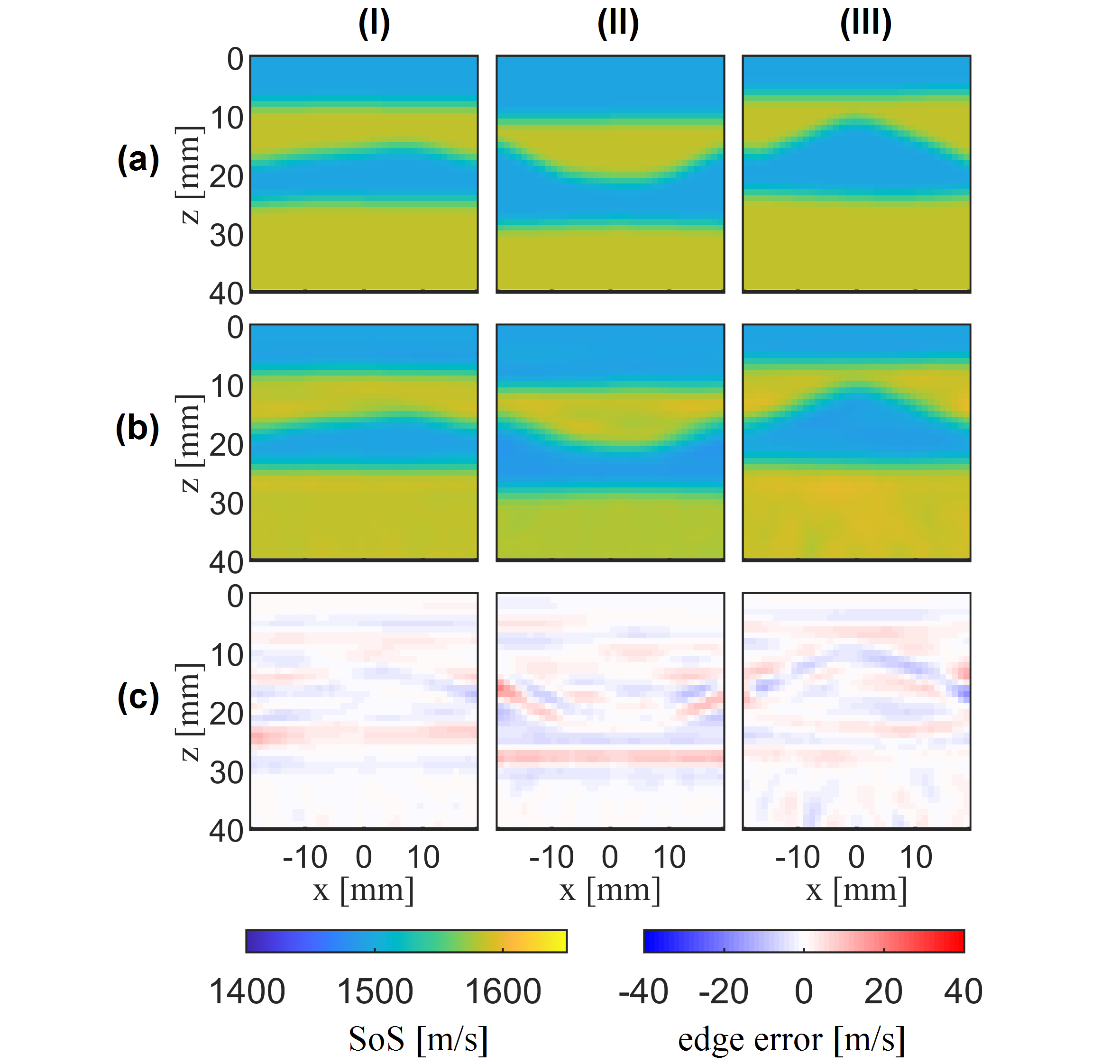}}
\caption{LI results (training set size $N=5000$) for the example tissue models (I) to (III) when using the wave model for ES simulation. (a) Target SoS maps. (b) Corresponding LI-SoS maps.(c) 2D-resolved edge error maps. } 
\label{fig: SoS maps full wave echo}
\end{figure}

Fig.~\ref{fig: SoS maps full wave echo} shows the LI-SoS maps for the three example tissue models. These results differ little from the ones obtained with LC, in agreement with the previous observations. This also holds for the edge errors: the 2D-resolved maps are similar to the ones for LC. This observation is confirmed by the RMS edge error distribution range in Fig.~\ref{fig: losses full wave echo}(b): $3.3\pm0.75$~m/s compared to the one for LC ($3.5\pm0.78$~m/s). 

\begin{figure}[!t]
\centerline{\includegraphics[width=\figurewidth]{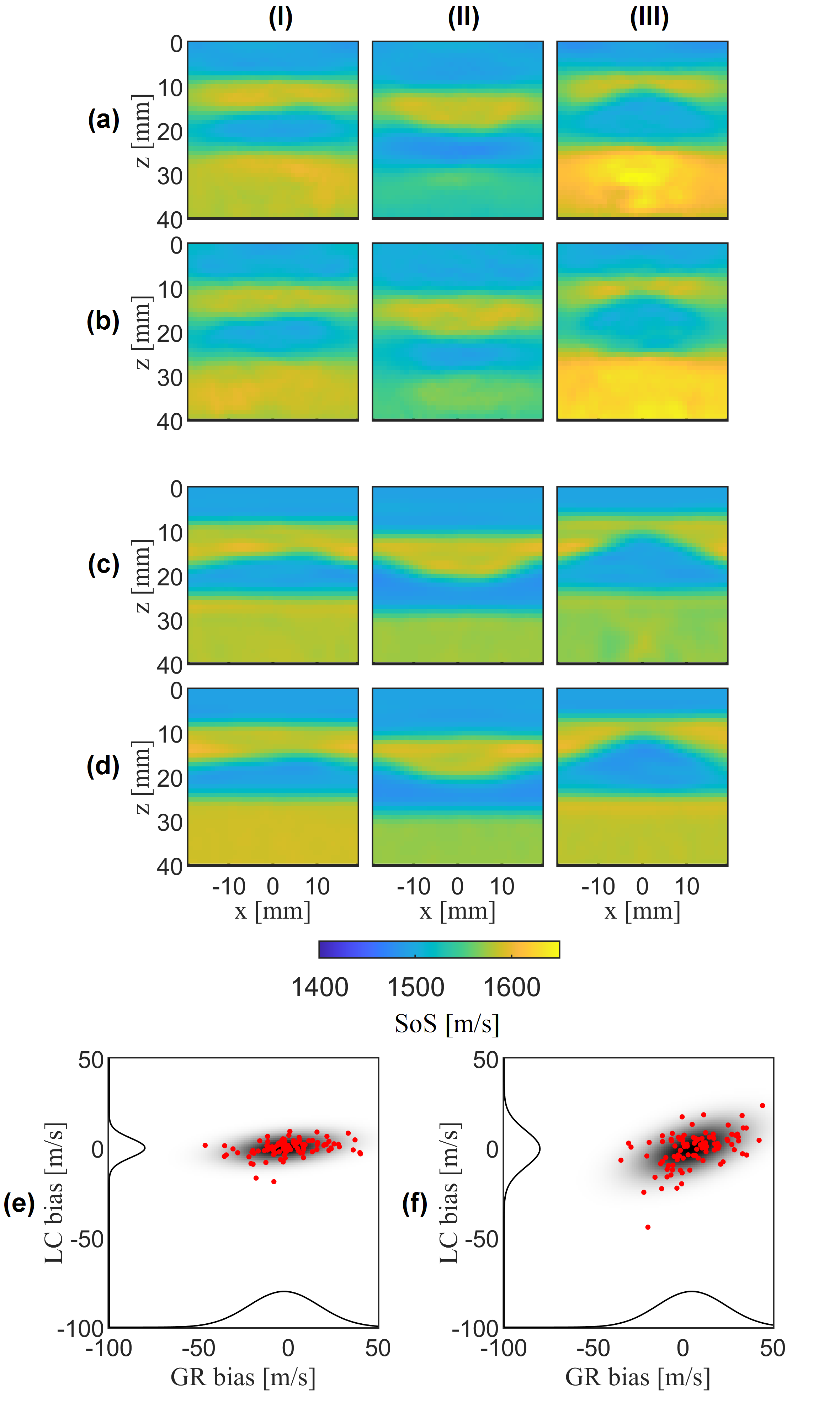}}
\caption{(a) to (d) Estimated SoS maps for the example tissue models (I) to (III) when departing from different beamforming SoS $c_0$, for GR (a: 1500~m/s, b: 1580~m/s) and for LC (c: 1500~m/s, b: 1580~m/s). (e)-(f) Layer L SoS biases obtained for the LC vs. GR for 1500~m/s and 1580~m/s, respectively. Normal
distribution fits (gray areas and profiles) are shown for 1000 test samples together with the bias values for 100 test samples (red dots).
} 
\label{fig: influence beamforming SoS}
\end{figure}

\subsection{Influence of beamforming speed of sound}
\label{sec: influence beamforming SoS} 
The beamforming SoS $c_0$ determines the offset of the true SoS relative to $c_0$ and thus the ES values. This dependence is ideally reversed in the reconstruction followed by Eq.~\ref{eq_slowness to SoS}, in a way that the final SoS map is independent of $c_0$. It has been hypothesized that---due to a velocity-depth ambiguity---the estimated SoS is biased depending on $c_0$~\cite{Ali2023ambiguity}. We have shown in the past that a bias towards $c_0$ can be explained by ES data degradation due to short-scale aberrations in heterogeneous tissue, but it is typically small in layered samples with piece-wise uniform SoS~\cite{jaeger2022pulse,jaeger2023heterogeneity}. In the present study, we have so far focused on $c_0$~=~$1540$~m/s. Fig.~\ref{fig: influence beamforming SoS} provides complementary results for $c_0$~=~$1500$~m/s and $1580$~m/s for the wave model. The GR-SoS maps of the example tissue models (I) to (III) confirm that the influence of $c_0$ is small. This observation is supported by the bias distributions: while a small shift of the distribution center is observed, it is much smaller than the change in $c_0$ (from $-2.4$~m/s to $4.7$~m/s between $1500$~m/s and $1580$~m/s) and than the distribution width (STD: $19$~m/s on average). The LC-SoS maps also show little influence of $c_0$, however, the bias distributions indicate that bias reduction by LC is more efficient for $c_0$~=~$1500$~m/s (STD: $5.4$~m/s, similar to what we observed for $1540$~m/s), than for $1580$~m/s (STD: $10$~m/s).  

\subsection{Physical phantom results}
\label{sec: physical results} 
\begin{figure}[!t]
\centerline{\includegraphics[width=\figurewidth]{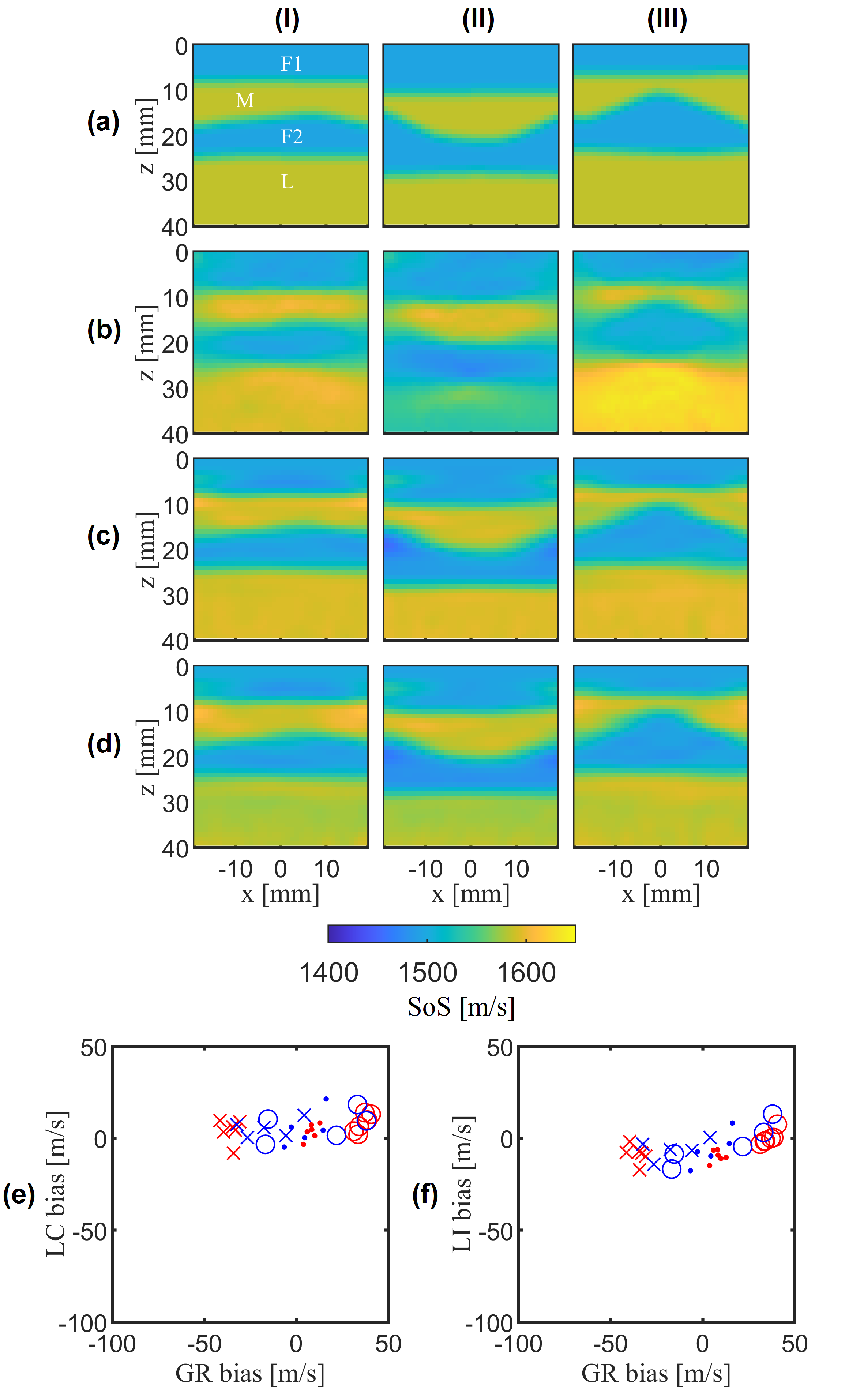}}
\caption{Physical twin results of GR and LC/LI approaches for the three example tissue models (I) to (III). (a) Target SoS maps derived from the known mold geometry and reference SoS measurements. (b) GR-SoS maps. (c) LC-SoS maps. (d) LI-SoS maps. (e)-(f) Layer L SoS distribution for LC and LI, respectively, vs. GR., for phantom (I) (dots), (II) (crosses), and (III) (circles), for different elevation positions (red) and different scan positions (blue).} 
\label{fig: real data results}
\end{figure}

Our study was based on simulations of the large number of training samples required for a good validation performance. It is of interest to see how well the simulation-trained linear operators perform on real data. For this purpose, we built physical twins of the example tissue models (I) to (III), from layers made of gelatine (25wt\% in water) and agar (2wt\% in water), with flour (2wt\%) for echogenicity. The geometry of the layers was determined by casting molds that were 3D-printed according to the simulated geometries. The SoS values of the gelatine and agar layers were determined by through-transmission measurements to $1585\pm5$~m/s and $1495\pm5$~m/s, respectively. 
The target and reconstructed SoS maps are shown in Figs.~\ref{fig: real data results}(a)-(d). We can make similar observations as for the simulations:
(i) With GR, the layer L's SoS is substantially biased depending on the layer geometry. While it is roughly correct for (I), it is underestimated for (II) and overestimated for (III). (ii) The layer L's SoS values with LC/LI show notably less geometry-dependent bias, seen as a more consistent value between the different phantoms than with GR. For each of the geometries (I) to (III), we collected 6 independent acquisitions by placing the imaging plane at different elevation positions. Fig.~\ref{fig: real data results}(e) and (f) show the reproducibility of the resulting layer L's biases, for LC and LI vs. GR, respectively (red markers). The mean STD values for these geometries are $3.6$~m/s, $5.3$~m/s, and $4.1$~m/s for GR, LC, and LI, respectively. The length of the phantoms along the imaging plane was $8$~cm. This allowed us to effectively create additional layer geometries by collecting 5 acquisitions while scanning the phantoms parallel to the imaging plane in steps of about $1$~cm. Fig.~\ref{fig: real data results}(e) and (f) show the layer L's bias values (blue marks). The distribution of the bias values for the total of 33 acquisitions ($3 \times (5+6)$) allows to appreciate the effect of LC/LI: while the center of the distribution differs between the two methods ($5.6$~m/s for LC vs. $-5.5$~m/s for LI), the STD for GR, LC, and LI  ($26$~m/s, $6.3$~m/s, and $7.2$~m/s, respectively) are in good agreement with the simulations. For these results, we used the $\mathbf{\Gamma}_\mathrm{opt}$ trained with $N$~=~$10000$. For $N$~=~$50000$, performance was slightly worse, with bias STD $7.3$~m/s (LC) and $7.9$~m/s (LI).  

\begin{table}[ht!]
\centering
\caption{Quantitative Metric Summary}
\label{tab:metric_table}
\renewcommand{\arraystretch}{1.2}
\footnotesize
\begin{tabular}{|l|c|c|c|c|}
\hline
\textbf{Scenario} & \textbf{N } & \textbf{Loss} & \textbf{Bias} & \textbf{RMS edge error}  \\ 
\hline

Linear GR & \multicolumn{1}{c|}{---} & \cellcolor{red!25}$4.2\text{e-}5$ & \cellcolor{blue!25}$-2.7\pm18$ & \cellcolor{green!25}$5.4\pm1.2$  \\ 
\hline

Wave GR   & \multicolumn{1}{c|}{---}  & \cellcolor{red!25}$5.9\text{e-}5$ & \cellcolor{blue!25}$-4.4\pm19$  & \cellcolor{green!25}$7.1\pm1.5$  \\ 
\hline

GR 1500  & \multicolumn{1}{c|}{---} & \cellcolor{red!25}--- & \cellcolor{blue!25}$-2.4\pm20$ & \cellcolor{green!25}---  \\
\hline

GR 1580   & \multicolumn{1}{c|}{---} & \cellcolor{red!25}---& \cellcolor{blue!25}$4.7\pm19$  & \cellcolor{green!25}---  \\ 
\hline

Phys. GR& \multicolumn{1}{c|}{---}  & \cellcolor{red!25}--- & \cellcolor{blue!25}$1.7\pm26$  & \cellcolor{green!25}---   \\ 
\hline

\multirow{3}{*}{Linear LC} 
    & $50$  & \cellcolor{red!25}$1.3\text{e-}5$ & \cellcolor{blue!25}$0.38\pm6.9$ & \cellcolor{green!25}--- \\ 
   & $1500$  & \cellcolor{red!25}$1.5\text{e-}6$ & \cellcolor{blue!25}$0.13\pm1.8$ & \cellcolor{green!25}--- \\
    & $10000$ & \cellcolor{red!25}$6.4\text{e-}7$ & \cellcolor{blue!25}$-0.06\pm1.2$ & \cellcolor{green!25}--- \\ 
    & $50000$ & \cellcolor{red!25}$5.4\text{e-}7$ & \cellcolor{blue!25}$-0.05\pm1.1$ & \cellcolor{green!25}$0.68 \pm 0.55$ \\ 
\hline

\multirow{4}{*}{Wave LC} 
    & $50$ & \cellcolor{red!25}$2.8\text{e-}5$ & \cellcolor{blue!25}$0.43\pm14$ & \cellcolor{green!25}---  \\ 
    & $1500$ & \cellcolor{red!25}$9.1\text{e-}6$ & \cellcolor{blue!25}$0.38\pm6.7$ & \cellcolor{green!25}---  \\
    & $10000$ & \cellcolor{red!25}$8.1\text{e-}6$ & \cellcolor{blue!25}$0.40\pm6.3$ & \cellcolor{green!25}---  \\ 
    & $50000$ & \cellcolor{red!25}$7.6\text{e-}6$ & \cellcolor{blue!25}$0.24\pm5.9$ & \cellcolor{green!25}$3.5 \pm 0.78$  \\ 
\hline

\multirow{3}{*}{Wave LI} 
    & $50$ & \cellcolor{red!25}$5.2\text{e-}5$ & \cellcolor{blue!25}$-3.0\pm19$ & \cellcolor{green!25}---  \\ 
    & $1500$ & \cellcolor{red!25}$1.1\text{e-}5$ & \cellcolor{blue!25}$0.21\pm7.6$ & \cellcolor{green!25}---  \\ 
   & $10000$ & \cellcolor{red!25}$8.0\text{e-}6$ & \cellcolor{blue!25}$0.40\pm6.4$ & \cellcolor{green!25}---  \\ 
    & $50000$ & \cellcolor{red!25}$6.7\text{e-}6$ & \cellcolor{blue!25}$0.17\pm5.6$ & \cellcolor{green!25}$3.3 \pm 0.75$  \\ 
\hline

\multirow{1}{*}{LC 1500 } 
    & $50000$ & \cellcolor{red!25}--- & \cellcolor{blue!25}$0.12\pm5.4$ & \cellcolor{green!25}---  \\ 
\hline

\multirow{1}{*}{LC 1580 } 
    & $50000$ & \cellcolor{red!25}--- & \cellcolor{blue!25}$-0.35\pm10$ & \cellcolor{green!25}---  \\ 
\hline
\multirow{1}{*}{Phys. LC} 
   & $10000$& \cellcolor{red!25}--- & \cellcolor{blue!25}$5.6\pm6.3$ & \cellcolor{green!25}---\\ 
\hline
\multirow{1}{*}{Phys. LI} 
    &$10000$ & \cellcolor{red!25}--- & \cellcolor{blue!25}$-5.5\pm7.2$ & \cellcolor{green!25}--- \\ 
\hline
\end{tabular}
\end{table}

\section{Discussion and Conclusion} 
\label{sec:discussion}
Our simulation study shows that the learning-based approaches to linear SoS inversion (LC and LI) yield substantially more accurate SoS maps than the pseudo-inverse with gradient regularization (GR). This is seen as a reduced validation loss value, accompanied by a substantial reduction of SoS biases in the liver-mimicking layer L. Very importantly, we have demonstrated that this bias reduction is achieved without compromising but even improving structural information. 
This observation is independent of whether ES is described by the straight-ray approximation (linear model) or by the full image generation process (wave model). 

For the linear model, the learning-based approach reduces the bias STD from $18$~m/s (GR) to $1.1$~m/s (LC). Thereby, the reconstructed SoS maps are an accurate representations of the target maps. Thus, the biases observed with GR are not an imperative result of an ill-posed inverse problem but are related to the smoothness constraint, which apparently does not capture the sample distribution well. 

For the wave model, we have shown that layer L' SoS biases with LC/LI are distributed within a larger range in agreement with an increased minimum loss compared to the linear model. These biases indicate a deviation of the relation between SD and ES from a linear behavior, which cannot be learned by a matrix operator. This non-linearity can emerge from misplacement of echoes---and thus of SD values---due to refraction and beamforming errors, and from a modification of the relation between SD and ES due to self-interference of propagating waves. The wave model contains these effects, and our results suggest that the precision limit for determining liver SoS with LI is $5.9$~m/s, three times smaller than the $19$~m/s with GR. 

The beamforming SoS $c_0$ is an important parameter determining the final outcome. In agreement with previous findings~\cite{stahli2020improved,jaeger2022pulse,jaeger2023heterogeneity}, the influence of $c_0$ on estimated SoS is small for layered structures with piece-wise uniform SoS, confirming that the observed biases are not caused by a velocity-depth ambiguity. Even though LC reduced the biases for all tested $c_0$, the performance is worse for $c_0$~=~$1580$~m/s than for $1540$~m/s and $1500$~m/s. The latter may again be explained by a non-linear relation between true SD and ES data: while the $1540$~m/s and $1500$~m/s are well inside the range of simulated tissue SoS ($1440$~m/s to $1615$~m/s), the $1580$~m/s are near the upper limit. The average SD is thus largest for this $c_0$ so that nonlinearities are expected to be most pronounced. 

We have proposed two ways of learning-based SoS reconstruction: either by defining (i) a matrix operator that corrects the already reconstructed GR-SD maps or (ii) a pseudo-inverse operator that acts directly on the ES data and thus replaces the GR pseudo-inverse (LI). We have shown that LI can perform better on test data than LC, provided that the training set size is sufficiently large. A disadvantage of the ES-based LI compared to the SD-based LC is that an ES data set contains more elements than a GR-SD map (in our case, 10 ES maps per GR-SD map), and thus the computation of $\mathbf{\Gamma}_\mathrm{opt}$ in~\eqref{eq_MR} is more expensive. Note, however, that the different sizes of $\mathbf{\Gamma}_\mathrm{opt}$ for the different methods, once computed, is not a limitation to real-time application: both approaches yield the same final matrix size as for GR (in LC, the product of $\mathbf{\Gamma}_\mathrm{opt}$ with $\mathbf{M}^\dag$ can be pre-computed). 

The phantom results confirm the efficacy of LC/LI in physical data: the bias STD was reduced from GR to LC/LI by a factor $4.1$ (LC) and $3.6$ (LI). However, unlike in the simulated test set, the bias distribution centers differ substantially between LC and LI, LI shows no benefit over LC, and increasing $N$ from $10000$ to $50000$ worsens performance. Such differences are not unexpected as the physical system can differ from the simulation model. To improve performance on physical data, one can envisage a transfer learning strategy \cite{obst2022transferlearning} where the simulation-trained $\mathbf{\Gamma}_\mathrm{opt}$ are fine-tuned on a limited number of phantoms.

\begin{figure*}[!ht]
\centerline{\includegraphics[width=\textwidth]{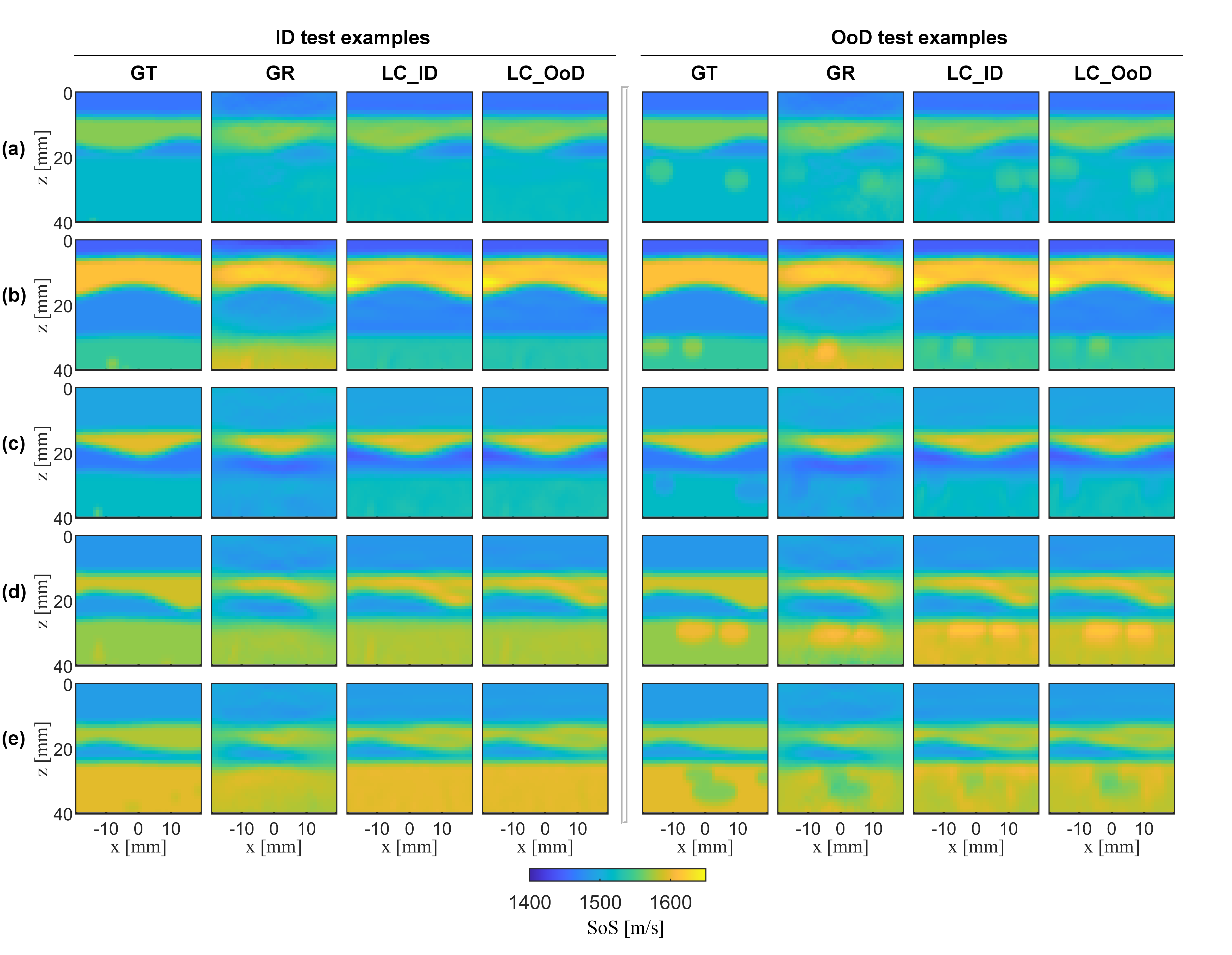}}
\caption{Illustration of LC performance on test samples without (in-distribution, ID) (left column) and with layer L inclusions (out-of-distribution, OoD) (right column). GT: ground truth; GR: GR-SoS maps; $\mathrm{LC_{ID}}$: LC-SoS maps when the training set includes only ID samples; $\mathrm{LC_{OoD}}$: LC-SoS maps when the training set includes OoD samples. (a) Example where no biases are seen. (b) Example where GR shows a positive layer L bias that is corrected by $\mathrm{LC_{ID}}$ and $\mathrm{LC_{OoD}}$. (c) Example where GR shows a negative layer L bias that is corrected by $\mathrm{LC_{ID}}$ and $\mathrm{LC_{OoD}}$. (d) No bias is seen with GR, but $\mathrm{LC_{ID}}$ and $\mathrm{LC_{OoD}}$ show bias in the OoD case. (e) Little bias is seen with GR and $\mathrm{LC_{ID}}$ and $\mathrm{LC_{OoD}}$, but LC fails at outlining inclusion, which is improved with $\mathrm{LC_{OoD}}$.} 
\label{fig: out of distribution results}
\end{figure*}

Any kind of regularization per definition biases the solution towards some implicit or explicit prior and works best on samples from within the statistical distribution favored by this prior. The same holds for the LC/LI approach: training and test samples were drawn from the same statistical distribution (layers of piece-wise uniform SoS), so it is natural to expect a good test performance. Here, we investigate a plausible out-of-distribution scenario where the uniformity of layer L is broken by inclusions mimicking liver tumors. For this, we simulated the same tissue models, once as described in Section~\ref{sec:echo_simulation} (in-distribution, ID), and once where---instead of vessels--- $1$ to $3$ circular areas are randomly placed inside layer L, with either positive or negative $30$~m/s SoS contrast (out-of-distribution, OoD). Fig.~\ref{fig: out of distribution results} shows representative test samples that allow to compare the performance of LC in the ID and OoD cases. Even if the same training data is used as in previous sections (i.e., $50000$ ID samples), LC reduces biases while inclusion visibility is maintained to a large part. However, the bias STD is larger for OoD ($9.5$~m/s)  than for ID ($5.28$~m/s) test samples (exemplified by (d)), and in some examples the inclusions are less visible with LC than with GR ((d) and (e)). When augmenting the training data with OoD samples (exemplarily $12000$), LC performance on ID test samples is slightly reduced (bias STD $5.5$~m/s, and see (a) to (e)), while it is improved on OoD test samples: the bias STD is reduced to $7.5$~m/s, and inclusion visibility is partially improved (see exemplary (e)). 

Our main statement is that SoS biases obtained with GR are to a large extent a regularization artifact and can be substantially reduced with LC or LI. It was practical to show this in an idealized scenario where the SoS inside tissue compartments is constant. In real tissue, however, aberrations of US waves at short-scale SoS variations inside compartments lead to increased artifact level~\cite{jaeger2023heterogeneity}. To thoroughly evaluate the potential of LC/LI in vivo, further research is required into understanding and optimizing the influence of the simulation design. This is ideally combined with aberration correction to reduce artifacts.  

Our study focused on linear SoS reconstruction that can be implemented as a pre-computable matrix operator with its benefit for real-time imaging. In LC/LI, this operator is constrained by optimizing the SD prediction over a set of training samples. In this context, the improved performance of LC/LI over GR can be understood if interpreting LC/LI as a distribution-specific optimum regularization that also includes local statistics, compared to the unspecific smoothness constraint implemented in GR.  
Non-linear artificial neural network (ANN)-based SoS reconstruction (e.g. \cite{Feigin2020deeplearning,simson2024investigating,chen2024deeptimeshift}) requires a pass through multiple network layers (as opposed to a single linear operator) but opens the possibility to account for a nonlinear relation between SD and ES \cite{chen2024deeptimeshift}. Our results indicate that a superior performance of ANNs compared to GR may in part stem from the above-mentioned distribution-specific regularization feature rather than from the nonlinear character. 

Finally, our study focused on the specific scenario of quantifying liver SoS. Being able to accurately determine liver SoS is promising for the diagnosis and grading of fatty liver disease. For successful translations to other potential uses of CUTE (such as cancer diagnosis), the technique will need to be adapted: (i) different training data is needed that represent the distribution expected in the specific desired application, and (ii) a different training target (such as a metric of contrast or spatial resolution) may be better suited than the mean-square deviation loss.

\bibliography{references}

@article{obst2022transferlearning,
  title={Improved linear regression prediction by transfer learning},
  author={Obst, David and Ghattas, Badih and Claudel, Sandra and Cugliari, Jairo and Goude, Yannig and Oppenheim, Georges},
  journal={Computational Statistics \& Data Analysis},
  volume={174},
  pages={107499},
  year={2022},
  publisher={Elsevier}
}

@article{Flynn2017risk, 
    title={Ultrasound Tomography Evaluation of Breast Density: A Comparison With Noncontrast Magnetic Resonance Imaging}, 
    author={O'Flynn, EAM and Fromageau, J and Ledger, AE and Messa, A and D'Aquino, A and Schoemaker, MJ and Schmidt, M and Duric, N and Swerdlow, AJ and Bamber, JC}, 
    journal={Invest Radiol}, 
    year={2017}, 
    volume={52}, 
    number={6}, 
    pages={343-348}, 
    doi={10.1097/RLI.0000000000000347}
}

@article{Ali2023ambiguity, 
    title={Sound speed estimation for distributed aberration correction in laterally varying media}, 
    author={Ali, Rehman and Mitcham, Trevor M. and Singh, Melanie and Doyley, Marvin M. and Bouchard, Richard R. and Dahl, Jeremy J. and Duric, Nebojsa},
    journal={IEEE Transactions on Computational Imaging}, 
    year={2023},
    volume={9},
    number={},
    pages={367-382},
    doi={10.1109/TCI.2023.3261507}}

@article{li2009vivo,
  title={In vivo breast sound-speed imaging with ultrasound tomography},
  author={Li, Cuiping and Duric, Nebojsa and Littrup, Peter and Huang, Lianjie},
  journal={Ultrasound in medicine \& biology},
  volume={35},
  number={10},
  pages={1615--1628},
  year={2009},
  publisher={Elsevier}
}

@article{wiskin2012non,
  title={Non-linear inverse scattering: High resolution quantitative breast tissue tomography},
  author={Wiskin, James and Borup, David T and Johnson, SA and Berggren, M},
  journal={The Journal of the Acoustical Society of America},
  volume={131},
  number={5},
  pages={3802--3813},
  year={2012},
  publisher={AIP Publishing}
}

@article{sandhu2015frequency,
  title={Frequency domain ultrasound waveform tomography: breast imaging using a ring transducer},
  author={Sandhu, GY and Li, Cuiping and Roy, Olivier and Schmidt, S and Duric, Neb},
  journal={Physics in Medicine \& Biology},
  volume={60},
  number={14},
  pages={5381-5398},
  year={2015},
  publisher={IOP Publishing}
}

@article{ruby2019breast,
title = {{Breast Cancer Assessment With Pulse-Echo Speed of Sound Ultrasound From Intrinsic Tissue Reflections: Proof-of-Concept}},
journal = {Investigative Radiology},
volume = {54},
pages = {419--427},
year = {2019},
doi = {https://doi.org/10.1097/RLI.0000000000000553},
author = {L. Ruby and S. J. Sanabria and K. Martini and K. J. Dedes and D. Vorburger and E. Oezkan and T. Frauenfelder and O. Goksel and M. B. Rominger}}

@article{Duric2007breast,
author = {Duric, Nebojsa and Littrup, Peter and Poulo, Lou and Babkin, Alex and Pevzner, Roman and Holsapple, Earle and Rama, Olsi and Glide, Carri},
title = {{Detection of breast cancer with ultrasound tomography: First results with the Computed Ultrasound Risk Evaluation (CURE) prototype}},
journal = {Medical Physics},
volume = {34},
number = {2},
pages = {773-785},
doi = {https://doi.org/10.1118/1.2432161},
year = {2007}
}

@article{boozari2010evaluation,
  title={Evaluation of sound speed for detection of liver fibrosis: prospective comparison with transient dynamic elastography and histology},
  author={Boozari, Bita and Potthoff, Andrej and Mederacke, Ingmar and Hahn, Andreas and Reising, Ansgar and Rifai, Kinan and Wedemeyer, Heiner and Bahr, Matthias and Kubicka, Stefan and Manns, Michael and others},
  journal={Journal of Ultrasound in Medicine},
  volume={29},
  number={11},
  pages={1581--1588},
  year={2010},
  publisher={Wiley Online Library}
}

@article{imbault2017robust,
  title={Robust sound speed estimation for ultrasound-based hepatic steatosis assessment},
  author={Imbault, Marion and Faccinetto, Alex and Osmanski, Bruno-F{\'e}lix and Tissier, Antoine and Deffieux, Thomas and Gennisson, Jean-Luc and Vilgrain, Val{\'e}rie and Tanter, Micka{\"e}l},
  journal={Physics in Medicine \& Biology},
  volume={62},
  number={9},
  pages={3582–3598 },
  year={2017},
  publisher={IOP Publishing},
  doi = {10.1088/1361-6560/aa6226},
}

@article{burgio2019ultrasonic,
  title={Ultrasonic adaptive sound speed estimation for the diagnosis and quantification of hepatic steatosis: a pilot study},
  author={Dioguardi Burgio, Marco and Imbault, Marion and Ronot, Maxime and Faccinetto, Alex and Van Beers, Bernard E and Rautou, Pierre-Emmanuel and Castera, Laurent and Gennisson, Jean-Luc and Tanter, Mickael and Vilgrain, Val{\'e}rie},
  journal={Ultraschall in der Medizin-European Journal of Ultrasound},
  volume={40},
  number={6},
  pages={722--733},
  year={2019},
  publisher={{\copyright} Georg Thieme Verlag KG},
  doi = {10.1055/a-0660-9465}
}

@article{telichko2022noninvasive,
doi = {10.1088/1361-6560/ac4562},
year = {2022},
publisher = {IOP Publishing},
volume = {67},
number = {1},
pages = {015007},
author = {Arsenii V Telichko and Rehman Ali and Thurston Brevett and Huaijun Wang and Jose G Vilches-Moure and Sukumar U Kumar and Ramasamy Paulmurugan and Jeremy J Dahl},
title = {Noninvasive estimation of local speed of sound by pulse-echo ultrasound in a rat model of nonalcoholic fatty liver},
journal = {Physics in Medicine \& Biology}}

@article{robinson1982registration,
  title={Measurement of velocity of propagation from ultrasonic pulse-echo data},
  author={Robinson, DE and Chen, F and Wilson, LS},
  journal={Ultrasound in medicine \& biology},
  volume={8},
  number={4},
  pages={413--420},
  year={1982},
  publisher={Elsevier}
}

@article{kondo1990crossed,
  title={An evaluation of an in vivo local sound speed estimation technique by the crossed beam method},
  author={Kondo, Masafumi and Takamizawa, Kinya and Hirama, Makoto and Okazaki, Kiyoshi and Iinuma, Kazuhiro and Takehara, Yasuaki},
  journal={Ultrasound in medicine \& biology},
  volume={16},
  number={1},
  pages={65--72},
  year={1990},
  publisher={Elsevier}
}

@article{hesse2013fullwave,
  title={Nonlinear simultaneous reconstruction of inhomogeneous compressibility and mass density distributions in unidirectional pulse-echo ultrasound imaging},
  author={Hesse, Markus C and Salehi, Leili and Schmitz, Georg},
  journal={Physics in Medicine \& Biology},
  volume={58},
  number={17},
  pages={6163},
  year={2013},
  publisher={IOP Publishing}
}

@article{Ali2022layeredMedia, 
    title={Local Sound Speed Estimation for Pulse-Echo Ultrasound in Layered Media}, 
    author={Ali, R and Telichko, AV and Wang, H and Sukumar, UK and Vilches-Moure, JG and Paulmurugan, R and Dahl, JJ}, 
    journal={IEEE Trans Ultrason Ferroelectr Freq Control}, 
    volume={69}, 
    number={2}, 
    pages={500-511}, 
    year={2022}, 
    doi={10.1109/TUFFC.2021.3124479}
}

@article{jakovljevic2018local,
  title={Local speed of sound estimation in tissue using pulse-echo ultrasound: Model-based approach},
  author={Jakovljevic, Marko and Hsieh, Scott and Ali, Rehman and Chau Loo Kung, Gustavo and Hyun, Dongwoon and Dahl, Jeremy J},
  journal={The Journal of the Acoustical Society of America},
  volume={144},
  number={1},
  pages={254--266},
  year={2018},
  publisher={AIP Publishing}
}

@article{sanabria2018spatial,
  title={Spatial domain reconstruction for imaging speed-of-sound with pulse-echo ultrasound: simulation and in vivo study},
  author={Sanabria, Sergio J and Ozkan, Ece and Rominger, Marga and Goksel, Orcun},
  journal={Physics in Medicine \& Biology},
  volume={63},
  number={21},
  pages={215015},
  year={2018},
  publisher={IOP Publishing}
}

@article{podkowa2020convolutional,
  title={The convolutional interpretation of registration-based plane wave steered pulse-echo local sound speed estimators},
  author={Podkowa, Anthony S and Oelze, Michael L},
  journal={Physics in Medicine \& Biology},
  volume={65},
  number={2},
  pages={025003},
  year={2020},
  publisher={IOP Publishing}
}

@ARTICLE{Feigin2020deeplearning,
  author={Feigin, Micha and Freedman, Daniel and Anthony, Brian W.},
  journal={IEEE Transactions on Biomedical Engineering}, 
  title={A Deep Learning Framework for Single-Sided Sound Speed Inversion in Medical Ultrasound}, 
  year={2020},
  volume={67},
  number={4},
  pages={1142-1151},
  doi={10.1109/TBME.2019.2931195}}

@article{heriard2023refraction,
  title={Refraction-based speed of sound estimation in layered media: an angular approach},
  author={H{\'e}riard-Dubreuil, Baptiste and Besson, Adrien and Wintzenrieth, Fr{\'e}d{\'e}ric and Cohen-Bacrie, Claude and Thiran, Jean-Philippe},
  journal={IEEE Transactions on Ultrasonics, Ferroelectrics, and Frequency Control},
  volume={70},
  number={6},
  pages={486--497},
  year={2023},
  publisher={IEEE}
}

@article{simson2024investigating,
title = {Investigating pulse-echo sound speed estimation in breast ultrasound with deep learning},
journal = {Ultrasonics},
volume = {137},
pages = {107179},
year = {2024},
issn = {0041-624X},
doi = {https://doi.org/10.1016/j.ultras.2023.107179},
author = {Walter A. Simson and Magdalini Paschali and Vasiliki Sideri-Lampretsa and Nassir Navab and Jeremy J. Dahl}}

@article{bezek2024learningforwardmodel,
  title={Learning the Imaging Model of Speed-of-Sound Reconstruction via a Convolutional Formulation},
  author={Bezek, Can Deniz and Haas, Maxim and Rau, Richard and Goksel, Orcun},
  journal={IEEE Transactions on Medical Imaging},
  year={2024},
  publisher={IEEE}
}

@article{chen2024deeptimeshift,
  title={Robust deep learning for pulse-echo speed of sound imaging via time-shift maps},
  author={Chen, Haotian and Han, Aiguo},
  journal={Authorea Preprints},
  year={2024},
  publisher={Authorea}
}

@article{jaeger2015computed,
  title={{Computed Ultrasound Tomography in Echo Mode for Imaging Speed of Sound Using Pulse-Echo Sonography: Proof of Principle}},
  author={Jaeger, Michael and Held, Gerrit and Peeters, Sara and Preisser, Stefan and Gr{\"u}nig, Michael and Frenz, Martin},
  journal={Ultrasound in Medicine \& Biology},
  volume={41},
  number={1},
  pages={235--250},
  year={2015},
  doi = {10.1016/j.ultrasmedbio.2014.05.019}
}

@INPROCEEDINGS{jaeger2023proceeding,
author={Jaeger, Michael and Gerber, Urs Richard and Yolgunlu, Parisa Salemi and Korta Martiartu, Naiara and Frenz, Martin},
booktitle={2023 IEEE International Ultrasonics Symposium (IUS)}, 
title={Learnt correction for regularization-related biases in pulse-echo speed-of-sound imaging}, 
year={2023},
volume={},
number={},
pages={1-4},
doi={10.1109/IUS51837.2023.10307408}}

@INPROCEEDINGS{jaeger2023heterogeneity, 
  author={Jaeger, Michael and Yolgunlu, Parisa Salemi and Korta Martiartu, Naiara and Frenz, Martin},
  booktitle={2023 IEEE International Ultrasonics Symposium (IUS)}, 
  title={Influence of tissue heterogeneity on the quantitative accuracy of pulse-echo speed-of-sound imaging: a simulation study}, 
  year={2023},
  volume={},
  number={},
  pages={1-4},
  doi={10.1109/IUS51837.2023.10308288}}

@article{stahli2023first,
title = {{First-in-human diagnostic study of hepatic steatosis with computed ultrasound tomography in echo mode}},
journal = {Communications Medicine},
volume = {3},
number = {1},
pages = {176},
year = {2023},
doi = {https://doi.org/10.1038/s43856-023-00409-3},
author = {P. St{\"a}hli and C. Becchetti and Korta Martiartu, N. and A. Berzigotti and M. Frenz and M. Jaeger}}

@article{stahli2020improved,
  title={Improved forward model for quantitative pulse-echo speed-of-sound imaging},
  author={St{\"a}hli, Patrick and Kuriakose, Maju and Frenz, Martin and Jaeger, Michel},
  journal={Ultrasonics},
  volume = {108},
  pages={106168},
  year={2020},
  doi = {10.1016/j.ultras.2020.106168}
}

@article{stahli2020bayesian,
  title={{Bayesian Approach for a Robust Speed-of-Sound Reconstruction Using Pulse-Echo Ultrasound}},
  author={St{\"a}hli, Patrick and Frenz, Martin and Jaeger, Michael},
  journal={IEEE Transactions on Medical Imaging}, 
  volume={40},
  number={2},
  pages={457--467},
  year={2021},
  doi={10.1109/TMI.2020.3029286}
}

@article{jaeger2022pulse,
title = {Pulse-echo speed-of-sound imaging using convex probes},
author = {Michael Jaeger and Patrick St\"ahli and Naiara {Korta Martiartu} and Parisa {Salemi Yolgunlu} and Thomas Frappart and Christophe Fraschini and Martin Frenz},
journal = {Physics in Medicine \& Biology},
doi = {10.1088/1361-6560/ac96c6},
year = {2022},
publisher = {IOP Publishing},
volume = {67},
number = {21},
pages = {215016}
}

@Article{salemi2023excluding,
AUTHOR = {Salemi Yolgunlu, Parisa and Korta Martiartu, Naiara and Gerber, Urs Richard and Frenz, Martin and Jaeger, Michael},
TITLE = {{Excluding Echo Shift Noise in Real-Time Pulse-Echo Speed-of-Sound Imaging}},
JOURNAL = {Sensors},
VOLUME = {23},
YEAR = {2023},
NUMBER = {12},
ARTICLE-NUMBER = {5598},
PubMedID = {37420762},
ISSN = {1424-8220},
DOI = {10.3390/s23125598}
}

@ARTICLE{montaldo2009coherent,
  author={Montaldo, Gabriel and Tanter, M. and Bercoff, J. and Benech, Nicolas and Fink, Mathias},
  journal={IEEE Transactions on Ultrasonics, Ferroelectrics, and Frequency Control},
  title={Coherent plane-wave compounding for very high frame rate ultrasonography and transient elastography}, 
  year={2009},
  volume={56},
  number={3},
  pages={489-506},
  doi={10.1109/TUFFC.2009.1067}}

@ARTICLE{Vyas2012has,
  author={Vyas, Urvi and Christensen, Douglas},
  journal={IEEE Transactions on Ultrasonics, Ferroelectrics, and Frequency Control}, 
  title={Ultrasound beam simulations in inhomogeneous tissue geometries using the hybrid angular spectrum method}, 
  year={2012},
  volume={59},
  number={6},
  pages={1093-1100},
  doi={10.1109/TUFFC.2012.2300}}

@ARTICLE{Beuret2024,
  author={Beuret, Samuel and H{\'e}riard-Dubreuil, Baptiste and Korta Martiartu, Naiara and Jaeger, Michael and Thiran, Jean-Philippe},
  journal={IEEE Transactions on Medical Imaging}, 
  title={{Windowed Radon Transform for Robust Speed-of-Sound Imaging With Pulse-Echo Ultrasound}},
  year={2024},
  volume={43},
  number={4},
  pages={1579-1593},
  doi={10.1109/TMI.2023.3343918}}

\end{document}